\documentclass[12pt,a4paper]{article}          
\usepackage{osajnl}
\usepackage[draft]{hyperref} 
\usepackage{epstopdf}
\usepackage{lipsum}
\usepackage{mathtools}
\usepackage{cuted}
\usepackage{graphicx}

\begin{document}

\title{Continuous-wave stability and multi-pulse structures in a universal Complex Ginzburg-Landau model for passively mode-locked lasers with saturable absorber}

\author{D\'efi Jr. Fandio Jubgang}
\affiliation{D\'epartement de Physique, Facult\'e des Sciences, Universit\'e de Sherbrooke, J1K-2R1 Sherbrooke Qu\'ebec, Canada}
\email{Defi.Junior.Jubgang.Fandio@USherbooke.ca}

\author{Alain M. Dikand\'e\footnote{Corresponding author}}
\affiliation{Laboratory of Research on Advanced Materials and Nonlinear Science (LaRAMaNS), Department of Physics, Faculty of Science, University of Buea P.O. Box 63 Buea, Cameroon}
\email{dikande.alain@ubuea.cm}

\author{A. Sunda-Meya}
\affiliation{Department of Physics and Computer Science, Xavier University of Louisiana, 1 Drexel Drive, New Orleans, Louisiana 70125, USA}
\email{asundame@xula.edu }

\begin{abstract}
The dynamics and stability of continuous-wave and multi-pulse structures are studied theoretically, for a generalized model of passively mode-locked fiber laser with an arbitrary nonlinearity. The model is characterized by a complex Ginzburg-Landau equation with saturable nonlinearity of a general form ($I^m/(1+\Gamma I)^n$), where $I$ is the field intensity, $m$ and $n$ are two positive real numbers and $\Gamma$ is the optical field saturation power. The analysis of fixed-point solutions of the governing equations, reveals an interesting loci of singular points in the amplitude-frequency plane consisting of zero, one or two fixed points depending upon the values of $m$ and $n$. The stability of continuous waves is analyzed within the framework of the modulational-instability theory, results demonstrate a bifurcation in the continuous-wave amplitude growth rate and propagation constant characteristic of multi-periodic wave structures. In the full nonlinear regime these multi-periodic wave structures turn out to be multi-pulse trains, unveiled via numerical simulations of the model nonlinear equation the rich variety of which is highlighted by considering different combinations of values for the pair ($m$,$n$). Results are consistent with previous analyses of the dynamics of multi-pulse structures in several contexts of passively mode-locked lasers with saturable absorber, as well as with predictions about the existence of multi-pulse structures and bound-state solitons in optical fibers with strong optical nonlinearity such as cubic-quintic and saturable nonlinearities. 
\end{abstract}


\maketitle

\section{Introduction}\label{section1}

Laser devices designed to generate ultrashort and high-power pulses have been the subject of a great deal of interest in the recent past, due to the enormous potential they offer for applications in a wide variety of modern communication technologies. \cite{1,2,3}. Among them, passively mode-locked fiber lasers have attracted much attention for they stand as the best ultrashort optical pulse sources available today. Indeed passively mode-locked fiber lasers possess a broad range of virtues as for instance their portability, simplicity and ability to generate transform-limited pulses up to subpicosecond regimes \cite{4}. Typically in a passively mode-locked fiber laser, high-intensity optical pulses are favored by the presence of a key device-usually a nonlinear element-characterized by an intensity-dependent response which promotes the formation of optical pulses over continuous waves (CWs) \cite{33}. This device acts like an optical amplifier and can be a saturable absorber such as a semiconductor saturable absorber \cite{44}, or an effective saturable absorber as for instance a nonlinear polarization switch \cite{66}, a nonlinear optical loop mirror \cite{77} or its variants \cite{99}. \\
Mode-locked fiber lasers with saturable absorber have been widely studied both experimentally and theoretically in the recent past \cite{a1,a2,a3,a4,a6,a8}. In standard setups where the optical gain medium is of weak nonlinearity, the propagation of the optical field can be approximated by a complex Ginzburg-Landau (CGL) equation with cubic nonlinearity \cite{a1,a6,a9,a10}. For this specific setup several theoretical studies have been carried out, the ultimate goal being to harvest as much insight as possible enabling a better understanding of peculiar features of the laser dynamics. These studies led to the relevant conclusion that despite its simplicity, the CGL equation accounts rather qualitatively well for the global dynamics of the laser system. In particular the cubic CGL equation provides fundamental insight onto the complexity of the system dynamics encompassing intrinsic properties of the system including dispersion,
self-phase modulation, cross-phase modulation, spectral filtering and gain and loss \cite{a1,a6,a9,a10}.\\ 
Relatively more recent studies of passively mode-locked lasers with saturable absorber, have established that optical gain media with very high nonlinear optical responses are more prone to the generation of stable high-intensity ultrashort pulse trains and multi-pulse structures including bound-state pulses \cite{p1,p2,p3}. Since the nonlinearity of the optical gain in this context cannot be of a Kerr type, several models have been proposed starting with the cubic-quintic CGL equation \cite{a11,a12,a13} as well as models with saturable nonlinearity (see e.g. ref. \cite{a8}). For classic semiconductor-doped
optical fiber media for instance, higher-order nonlinear terms beyond the cubic one are known to favor strongly nonlinear pulse structures, multiple-pulse and
bound-pulse structures as can be obtained both from direct numerical simulations of the CGL equations \cite{p1} and a modulational-instability analysis of CWs \cite{pd1,pd2}. \\
The advent of new passively mode-locked fiber laser setups, characterized by relatively stronger nonlinear optical gain media, calls to a need for proper modeling of their underlying properties. In this respect a model of CGL equation was recently introduced \cite{Dikande2017} to address the issue of self-starting for a general family of passively mode-locked fiber lasers with saturable absorber. In this model the nonlinearity is of the general form $I^m/(1+\Gamma I)^n$ where $I$ is the field intensity, and $m$, $n$ are real positive parameters. Besides its proposed application to passively mode-locked fiber lasers possessing relatively strong nonlinear gain media, this model is equally suitably adapted for the study of optical field propagation in exotic materials where optical nonlinear responses are enhanced \cite{Mikhail2014,Petmegni2017}. In fact this new model includes most of the existing forms of CGL equations proposed in the studies of passively mode-locked fiber lasers (e.g. the master equation considered in ref. \cite{a6} corresponds to $m = 2$ and $n = 0$, the master equation in ref. \cite{a8} corresponds to $m = 0$ and $n = 1$, the cubic-quintic CGL equation used in ref. \cite{a11} is recovered by expanding the nonlinear term to the second order for large optical field saturation power $\Gamma^{-1}$, with $m=2$ and $n$ arbitrary but nonzero, etc.), while providing rich perspectives for possible novel passively mode-locked laser setups with yet to be experimentally explored self-starting features, related to non-Kerr nonlinearities of their active media.\\
In general passively mode-locked lasers can sustain CW and pulse structures, depending both on the input power and on the competition between dispersion and nonlinearity. When the laser system operates in the anomalous dispersion regime, pulse and multi-pulse envelopes are often associated with the modulational instability of a steady-state CW. A small perturbation of the CW amplitude will grow upon propagation, such that nonlinear effects become gradually manifest and reshape the unstable CW field into pulses with soliton features. At higher perturbation order, a very rich plethora of multi-pulse structures can emerge including harmonically-mode-locked vector solitons \cite{Huang2016,Mesumbe2019}, bright-dark solitons \cite{Dikande2010,Dikande2011,MbiedaPetmegni2017}, soliton molecules \cite{Huang2016,brice}, elliptic-type soliton trains or soliton crystals \cite{Amrani2011,Jubgang2015,welak,fandioa}, soliton combs \cite{DikandeBitha2019,DikandeBitha2019a} and so on.\\
In this work we are interested in the dynamics of the family of fiber-laser systems which can be described by the model proposed in ref. \cite{Dikande2017}, with emphasis on CW and multi-pulse regimes of operation. To start we derive dynamical equations for the optical field amplitude and instantaneous frequency, and then examine characteristic features of their fixed points. A modulational-instability analysis of CWs reveals an increase in amplitude of small perturbation to the CW, for a specific range of the perturbation modulation frequency. The correlation between unstable CWs and laser self-starting is revised as a means to generate multi-pulse structures. Using a sixth-order Runge-Kutta scheme adapted from Luther \cite{luth}, numerical simulations of these multi-pulse structures are implemented for some physically relevant combinations of the real parameters $m$ and $n$.

\section{The model}\label{section2}
The propagation of an optical field in passively mode-locked lasers can exhibit several effects including dispersion, nonlinearity, linear and nonlinear gain and loss, saturable absorber effects and so on. In this study we are interested in a family of fiber lasers with distinct types of nonlinearity such as Kerr nonlinearity, nonlinear absorption and their saturation as observed in fiber-ring lasers \cite{Amrani2011,Jubgang2015}, figure-eight\cite{a6,Huang2016} and in some solid-state lasers \cite{a8}. For such systems, the governing equation for laser propagation can be represented by a CGL equation of the form \cite{Dikande2017}:
\begin{equation}
u_z =(g-\rho + i\theta)u+(C+i D)u_{tt}+\frac{(\gamma_r+ i \gamma_{im})}{(1+\Gamma|u|^2)^n}|u|^mu, \label{eq:one}
\end{equation}
where $u=u(z,t)$ is the optical field, $u_z=\partial u/\partial z$,  $u_{tt}=\partial^2 u/\partial t^2$, $z$ is the round-trip number and $t$ is the normalized time. Parameters $g$, $\rho$ and $\theta$  are the linear gain, loss (both assumed constant) and phase change over round-trip, respectively. $C$ and $D$ are respectively, the spectral filtering and the group-velocity dispersion of the gain medium. $\gamma_r$ and $\gamma_{im}$ are real and imaginary parts of the nonlinear coupling, and $\Gamma$ depicts the nonlinear saturation coefficient (inversely proportional to the field saturation power). \par 
In its present form eq. (\ref{eq:one}) requires $D>0$ for anomalous group-velocity dispersion, $\gamma_r<0$ and $\gamma_{im}>0$ to represent the real and imaginary parts of the nonlinear gain. Instructively $\gamma_r$ and $\gamma_{im}$ must be negative to guarantee self-starting for the family of mode-locked fiber lasers ($m=0,n\neq0$) \cite{a8}, in the anomalous regime of dispersion.\\
Most generally the quantities $m$ and $n$ are real and positive such that the couple ($m,n$) determines the specific type of nonlinearity prevailing in the active medium. Some physical previously discussed include: 
\begin{enumerate}
\item ($m,n$) = $(2,0)$, corresponding to the most common Kerr fiber lasers \cite{a6},
\item ($m,n$) = $(2,1)$, this corresponds to fiber lasers with active media like $CdS_{1-x}Se_x$-doped glasses that exhibit a second-order Kerr nonlinearity and a saturation field power \cite{a6,pd1},
\item ($m,n$) = $(0,1)$, corresponding to a model very close to the one discussed in ref. \cite{a8} for Cr:ZnSe and Ti:Sapphire lasers, 
\item ($m,n$) = $(2,2)$, this case is amenable, for relatively small nonlinearity-saturation coefficient, to the CGL equation with cubic-quintic nonlinearity \cite{a11}. 
\end{enumerate}
We will focus only on few combinations of numerical values of the pair ($m,n$), keeping in mind that any other combination involving real and positive values of the two paramaters is also physical.

\section{Stationary solutions}\label{section3}
\begin{figure}[bt]
\begin{center}
\includegraphics[width =0.45\columnwidth, height=1.7 in]{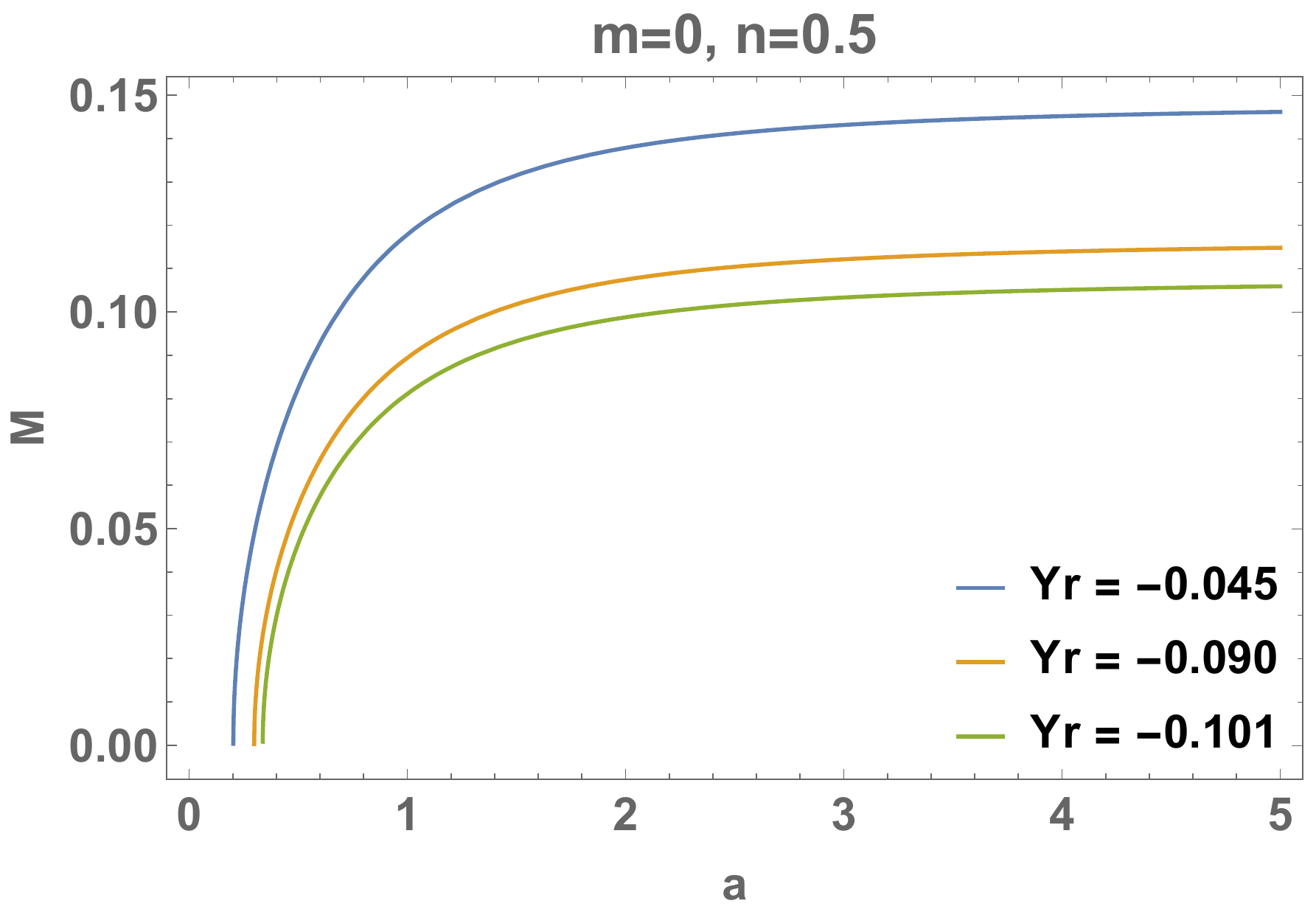}
\includegraphics[width =0.45\columnwidth, height=1.7 in]{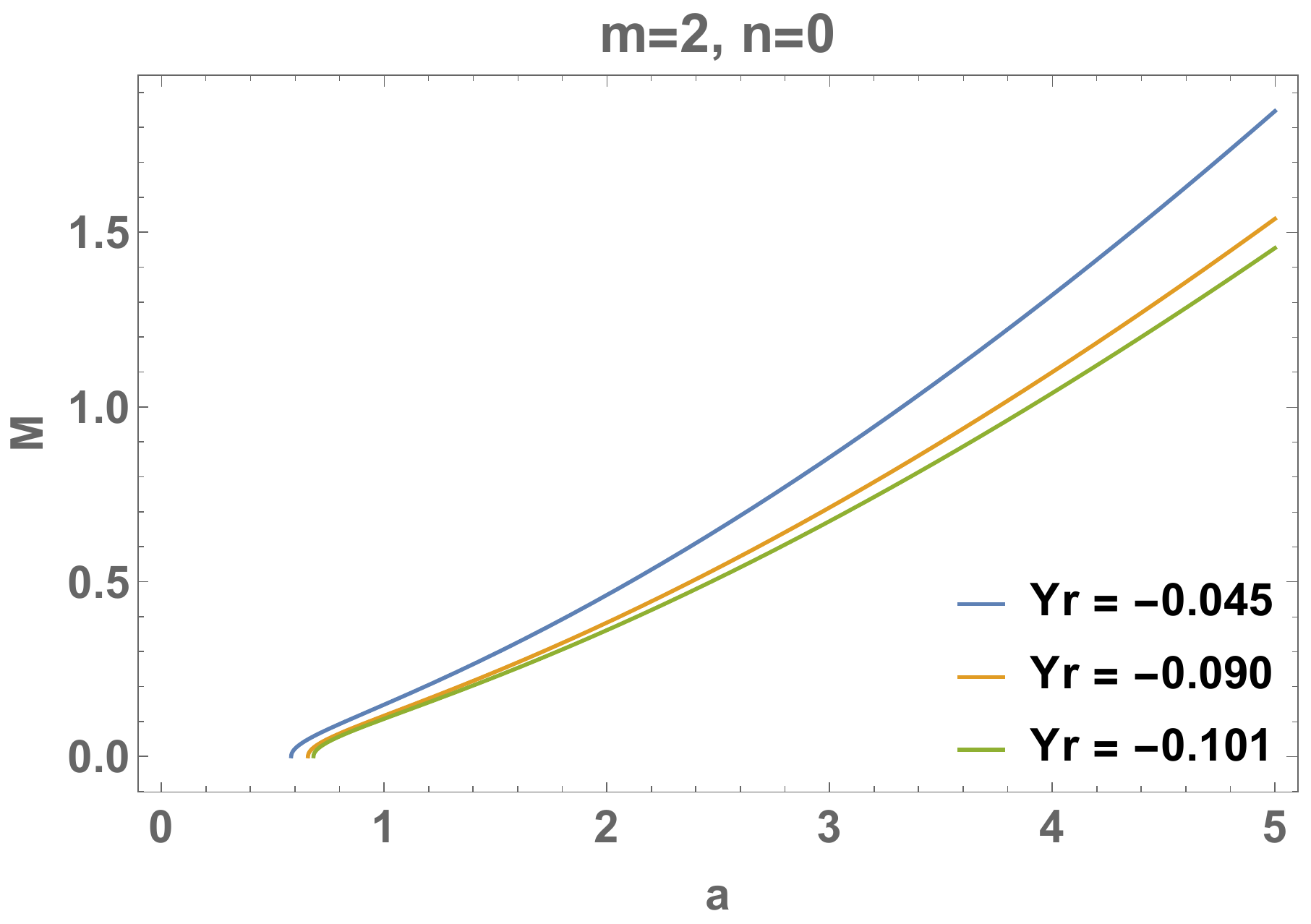}
\includegraphics[width =0.45\columnwidth, height=1.7 in]{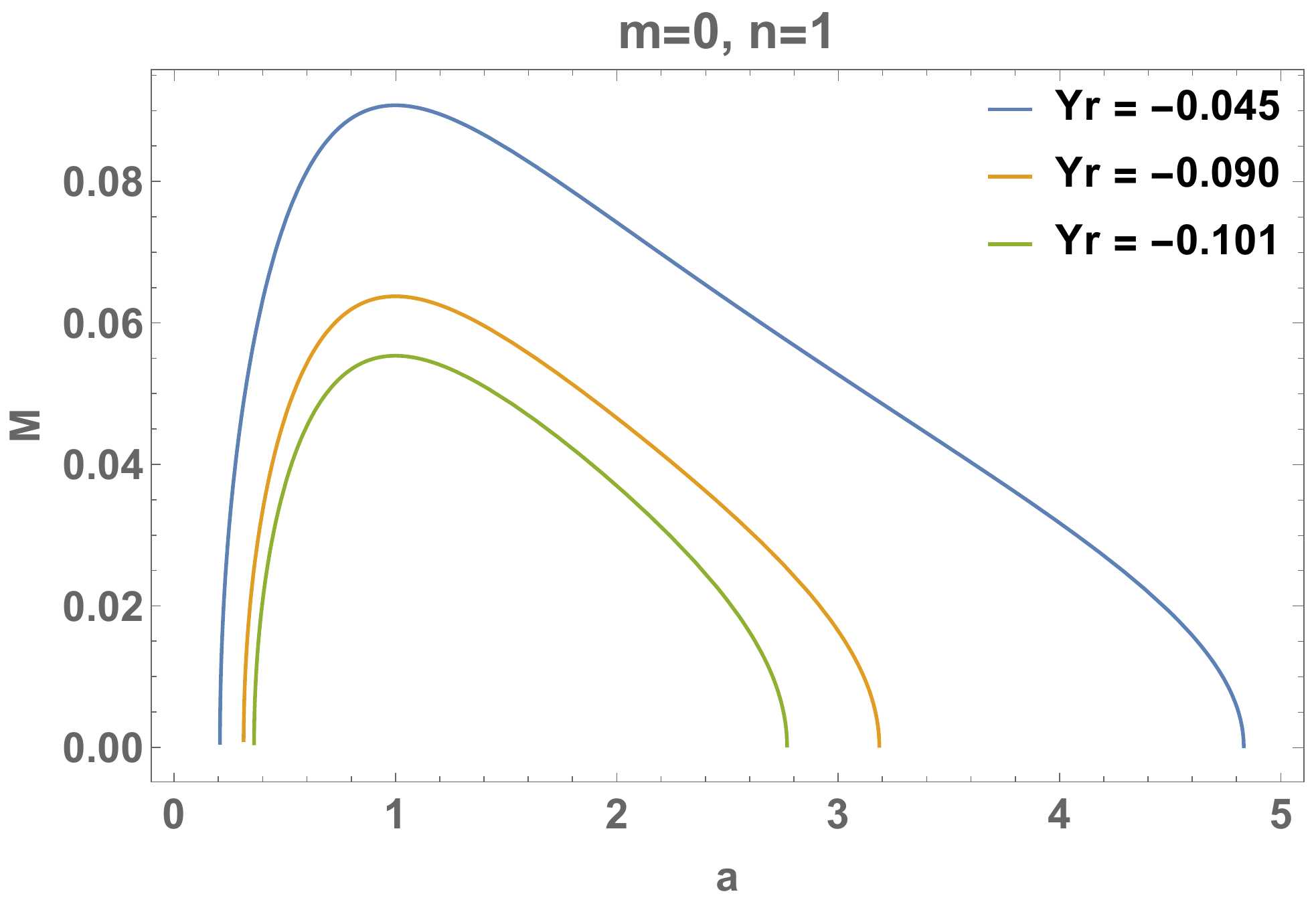}
\includegraphics[width =0.45\columnwidth, height=1.7 in]{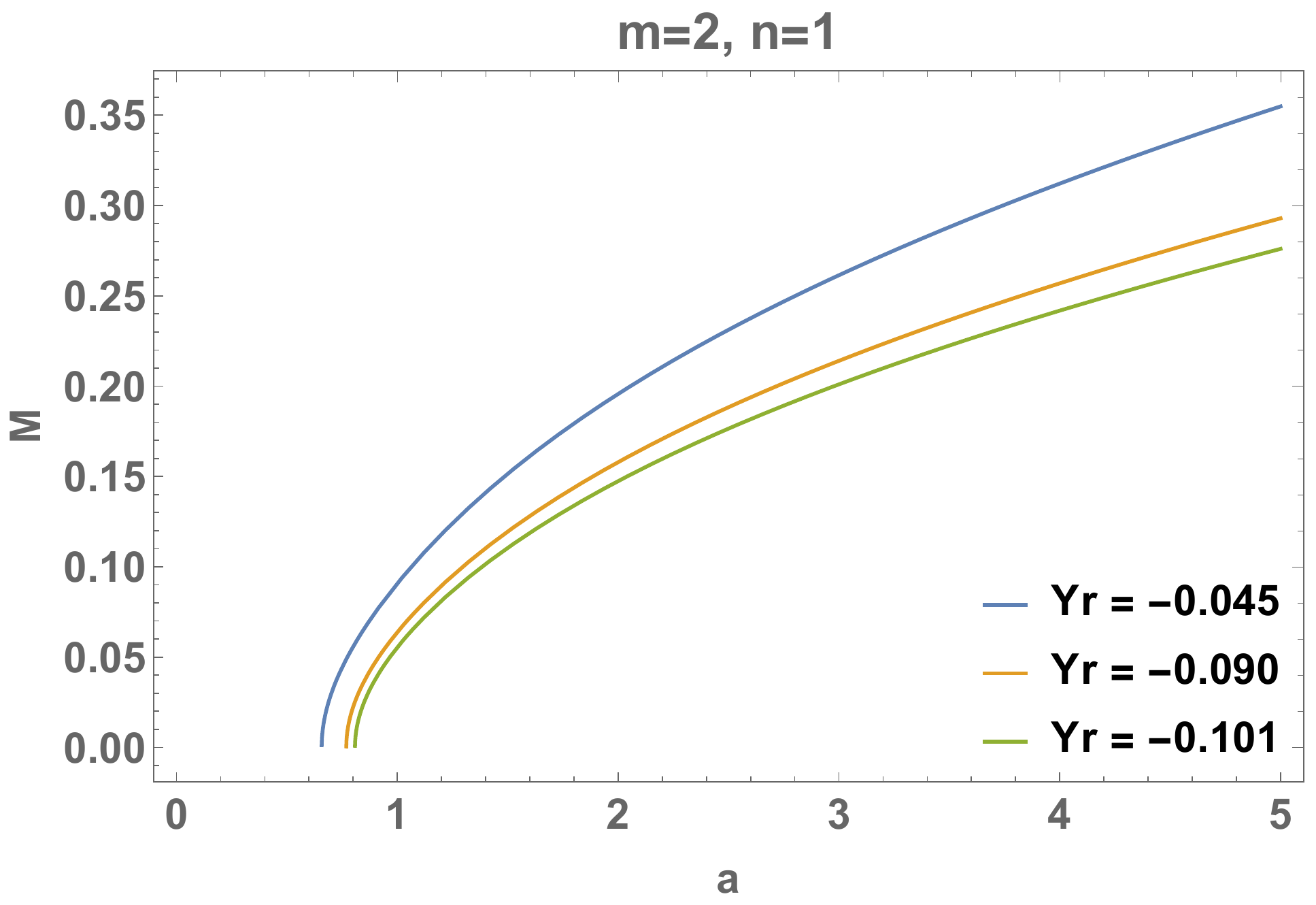}
\includegraphics[width =0.45\columnwidth, height=1.7 in]{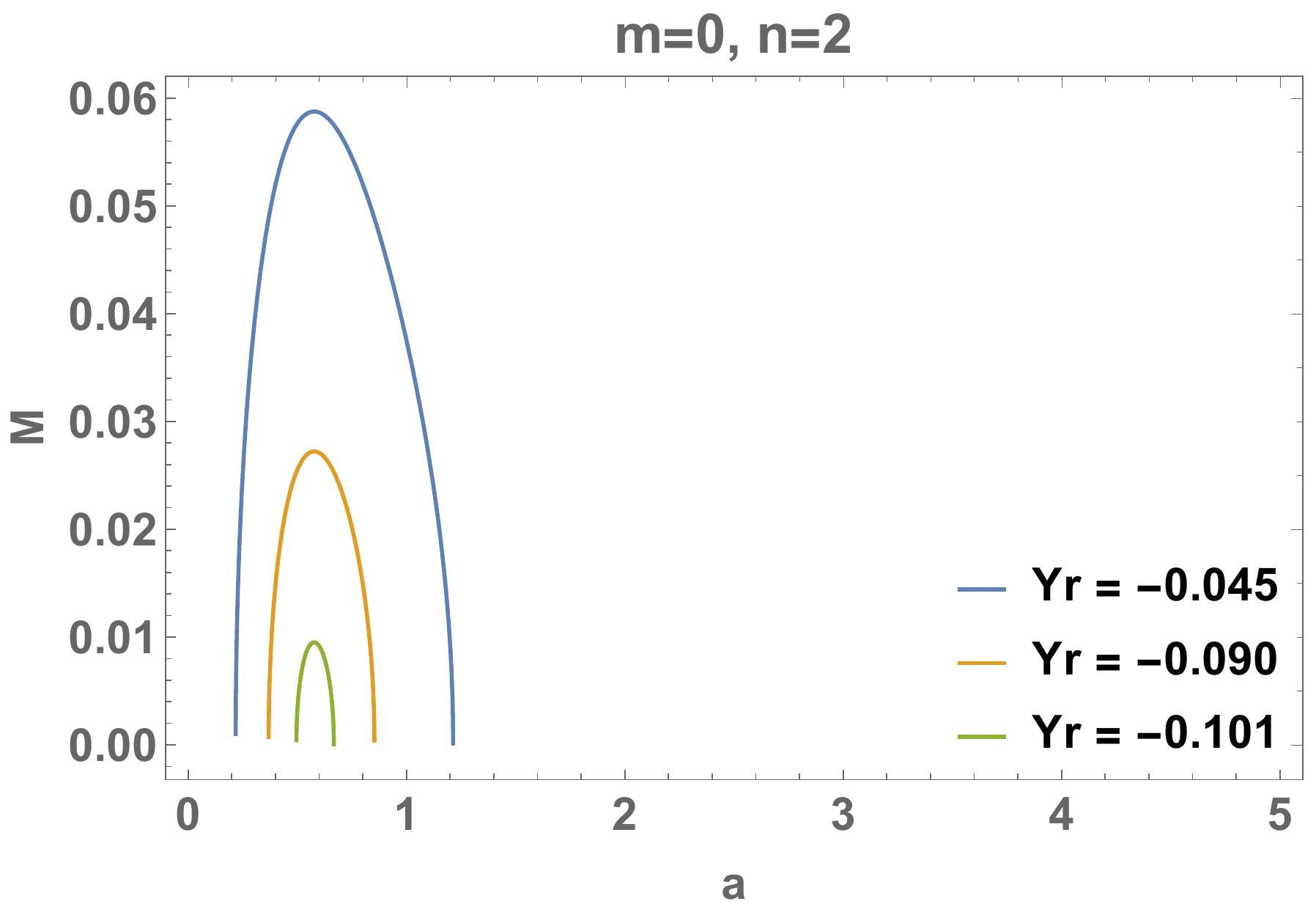}
\includegraphics[width =0.45\columnwidth, height=1.7 in]{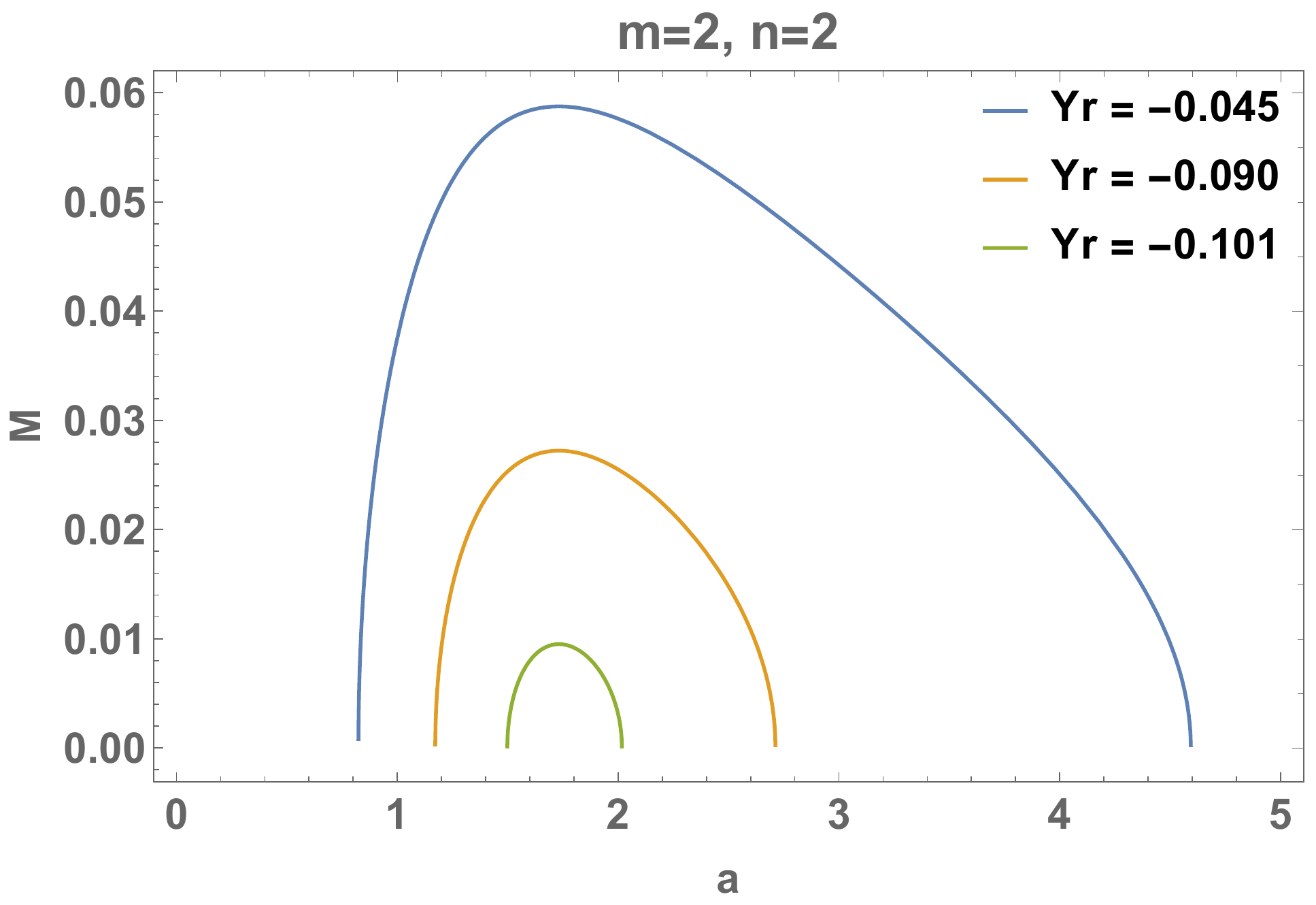}
\end{center}
\caption{(Color online) Loci of the stationary solutions to eq. (\ref{eq:fivea}) in the upper-half positive a-M plane, for different values of the pair ($m,n$). The solutions were obtained at different non-linear gain values $\gamma_{r}$ as indicate in the legends. Other parameters were fixed to $C=5$, $\gamma_{im}=1$, $\Gamma=1$, $\theta=-0.08$ and $g-\rho=0.09$} \label{fig:one}
\end{figure}

Eq. (\ref{eq:one}) admits both CW and pulse solutions. To find these solutions we formulate the optical propagating field in terms of the ansatz \cite{a11}: 
\begin{equation}
u(z,\tau)=a(\tau)\exp\bigg[ i \phi(\tau)-i \omega z\bigg], \label{eq:two}
\end{equation}
where $a$ and $\phi$ are real valued functions of the field amplitude and phase, respectively, and $\tau$= $t-vz$ is the new time variable with $v$ the inverse velocity and $\omega$ a nonlinear shift in the propagation constant \cite{a11}. Substituting eq. (\ref{eq:two}) into eq. (\ref{eq:one}), and separating the real and imaginary parts, the following set of coupled second-order nonlinear ordinary differential equations are derived for $a$ and $\phi$:
\begin{eqnarray}
C a_{\tau \tau} + (v-2 D \phi_\tau)a_\tau + (g-\rho - C \phi^2_\tau - D\phi_{\tau\tau}) a \nonumber \\
 + \frac{\gamma_r}{(1+\Gamma a^2)^n}a^{m+1} = 0 &,& \label{eq:threea}\\
D a_{\tau \tau} + 2 C \phi_{\tau}a_{\tau}+ (v \phi_\tau + \omega+\theta + C \phi_{\tau \tau} - D \phi^2_{\tau})a \nonumber \\  
+ \frac{\gamma_{im}}{(1+\Gamma a^2)^n} a^{m+1} = 0 &.& \label{eq:threeb}
\end{eqnarray}
Defining the instantaneous frequency as $M =\phi'$, and the field transient $y=a'$ where the primes refers to partial derivative with respect to $\tau$, eqs. (\ref{eq:threea})-(\ref{eq:threeb}) can be rewritten into a more convenient form as:

\begin{eqnarray}
y' &=& M^2 a - \frac{(Cvy+DvMa)}{C^2+D^2} - \frac{C(g-\rho)+D(\omega+\theta)}{C^2+D^2} a \nonumber \\
 &-& \frac{(C \gamma_r + D \gamma_{im})}{C^2+D^2} \frac{a^{m+1}}{(1+\Gamma a^2)^n}, \label{eq:foura}\\
M ' &=& \frac{D(g-\rho)-C(\omega+\theta)}{C^2+D^2} - \left(\frac{2 y}{a} + \frac{C v}{C^2+D^2}\right)M \nonumber \\ 
&+& \frac{D\gamma_r-C\gamma_{im}}{C^2+D^2} \frac{a^m}{(1+\Gamma a^2)^n},\\
a' &=& y,\label{eq:fourc} \\
\phi'&=&M. \label{eq:fourd}
\end{eqnarray}
Eqs. (\ref{eq:foura})-(\ref{eq:fourd}) are four coupled first-order nonlinear ordinary differential equations, which define the uniformly translating solutions to the generalized CGL equation eq. (\ref{eq:one}). For the rest of the study we consider only the context of mode-locked fiber lasers operating in the anomalous dispersion regime, such that $D = 1$ for simplicity. \\
Stationary solutions to eqs. (\ref{eq:foura})-(\ref{eq:fourd}) correspond to the loci ($a, M$), and determined by the fixed-point requirements i.e. $M'=0$, $y'=0$ and $y=0$. These equations can be rewritten with $\omega$ as a free parameter, in the following form:
\begin{eqnarray}
M^2 - \frac{C(g-\rho)-D(\omega+\theta)}{C^2+D^2} - \frac{C\gamma_r + D\gamma_{im}}{C^2+D^2}\frac{a^{m+1}}{(1+\Gamma a^2)^n} = 0, \label{eq:fivea}\\
(D\gamma_r-C\gamma_{im})a^m + (D(g-\rho)-C(\omega+\theta))(1+\Gamma a^2)^n = 0, \label{eq:fiveb}
\end{eqnarray}
when $v$ is assumed zero \cite{a11}. Eq. (\ref{eq:fivea}) admits a trivial solution ($a=0$, $M=0$) at the specific value of $\omega=C(g-\rho)/D-\theta$. The trajectory in the a-M plane that connects the origin with a nontrivial point corresponds to a front solution. When the trajectory connects two nontrivial points, the solution is either a sink or a source. \\
The amplitudes corresponding to the $M=0$ frequency are the flow derived by solving:
\begin{eqnarray}
[(C^2&-&D^2)\gamma_r+2DC\gamma_{im}] a^m \nonumber \\
&+& (C^2-D^2)(g-\rho)(1+\Gamma a^2)^n = 0, \label{eq:six}
\end{eqnarray}
where $\omega$ is eliminated in the equation by substituting eq. (\ref{eq:fivea}) into eq. (\ref{eq:fiveb}). Eq. (\ref{eq:six}) admits none or one root if either of $m$ or $n$ is zero, and two or more distinct roots when $m$ and $n$ are non-zero. However, it should be emphasized that some roots could be negative as for instance the case ($2,1$). Fig. \ref{fig:one} illustrates the stationary solutions obtained for different sets of ($m,n$), in the upper-right quadrant ($a>0$) of the a-M plane.\\
\begin{figure}[bt]
\begin{center}
\includegraphics[width =0.45\columnwidth, height=1.7 in]{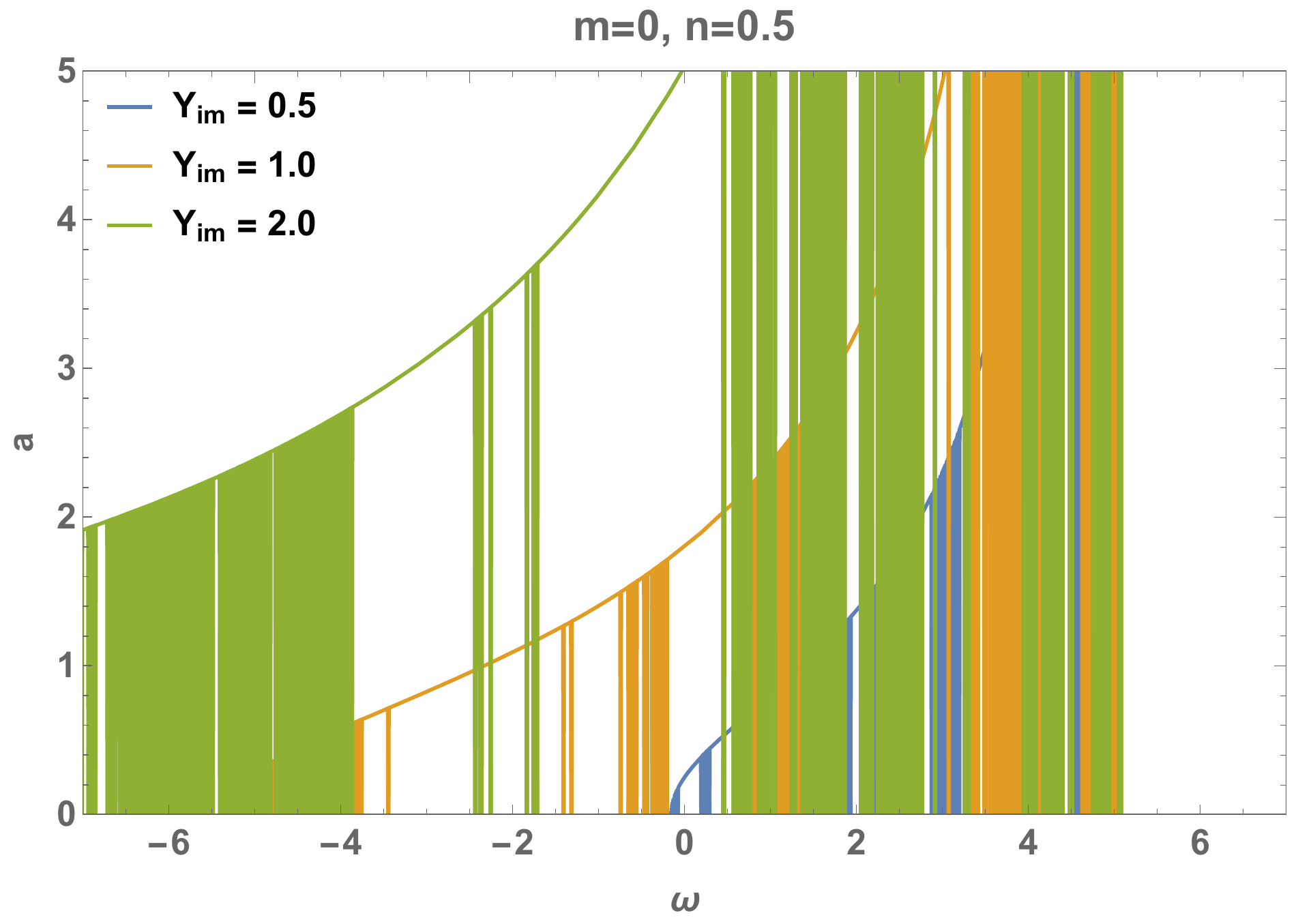}
\includegraphics[width =0.45\columnwidth, height=1.7 in]{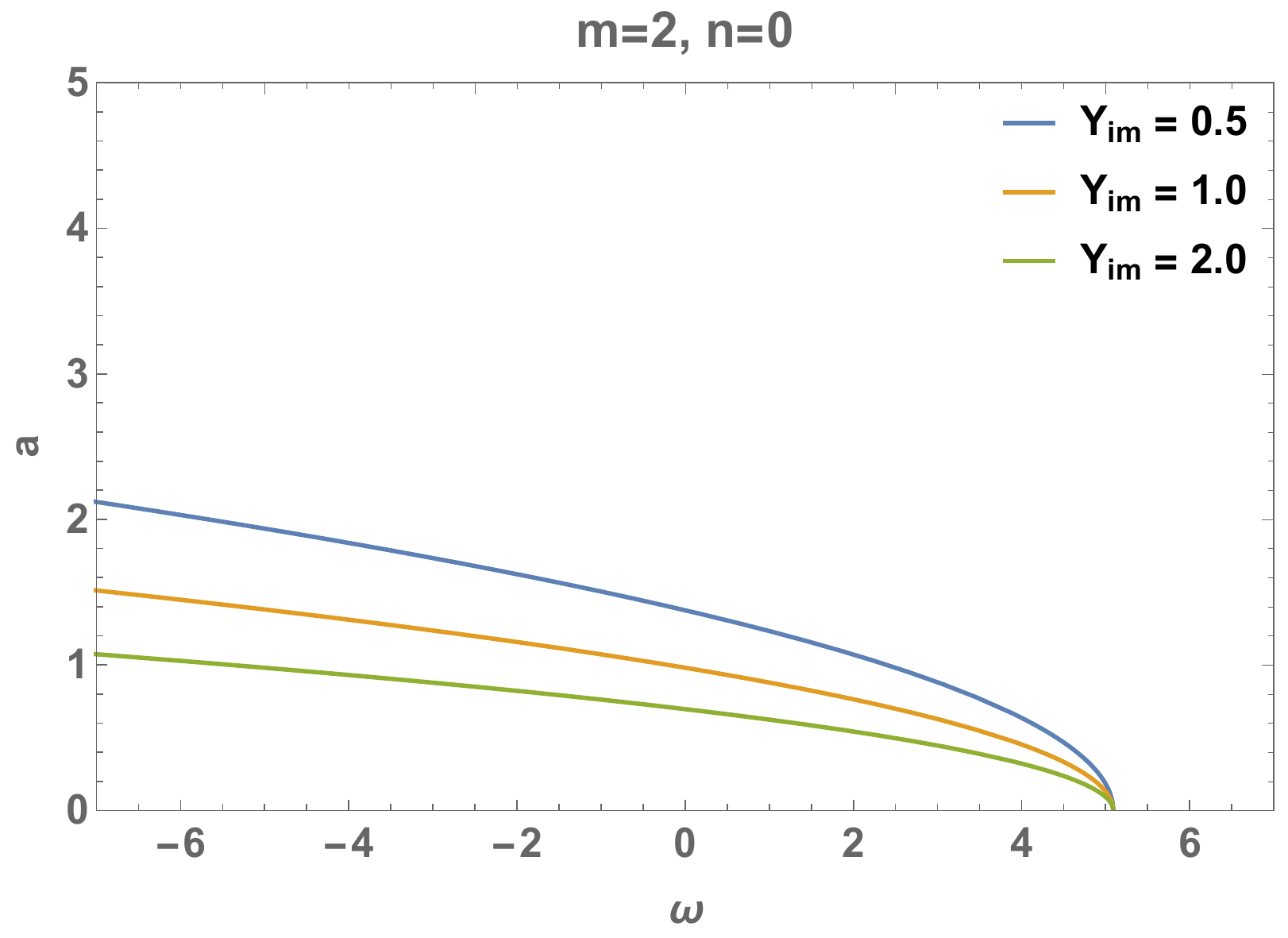}
\includegraphics[width =0.45\columnwidth, height=1.7 in]{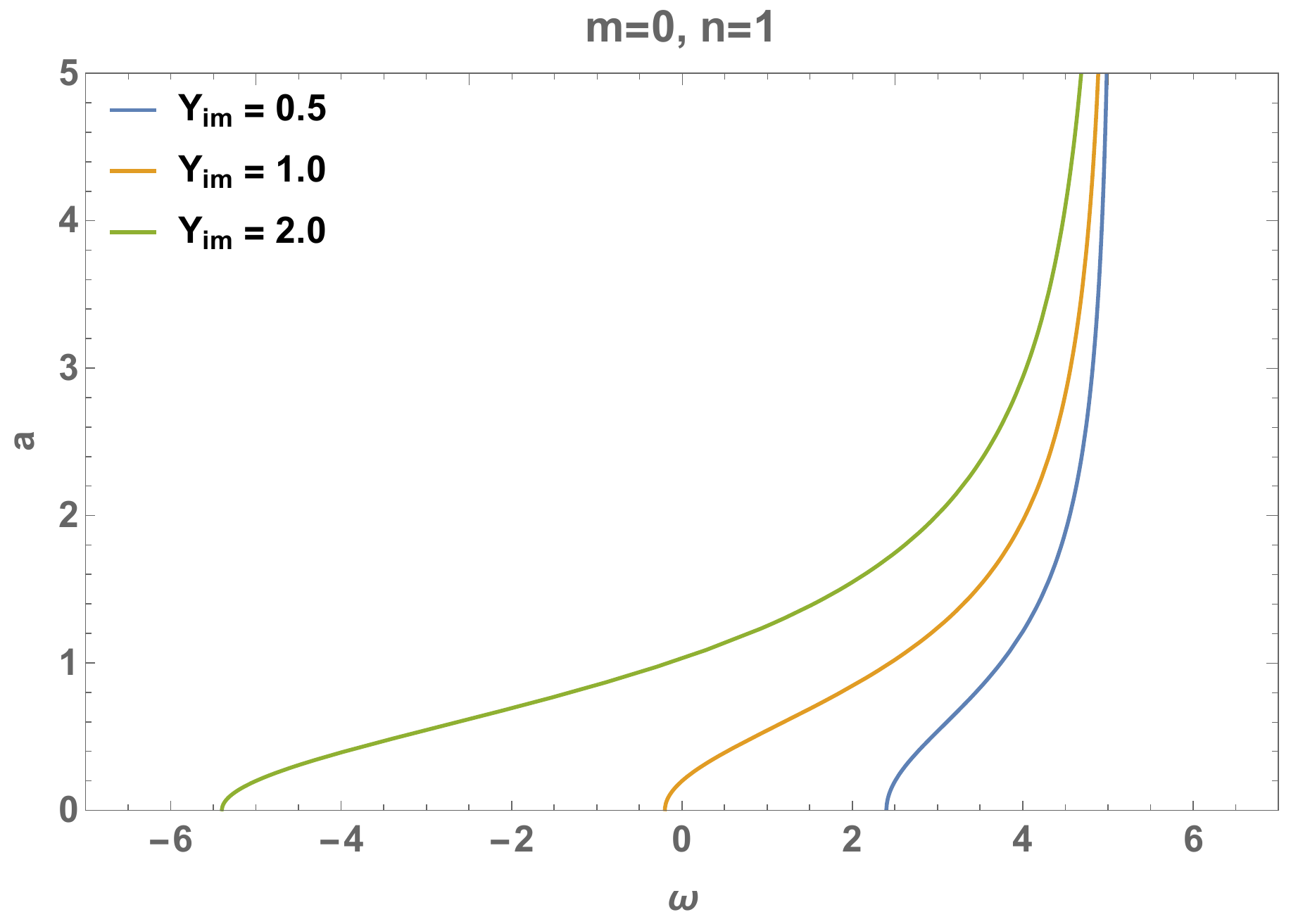}
\includegraphics[width =0.45\columnwidth, height=1.7 in]{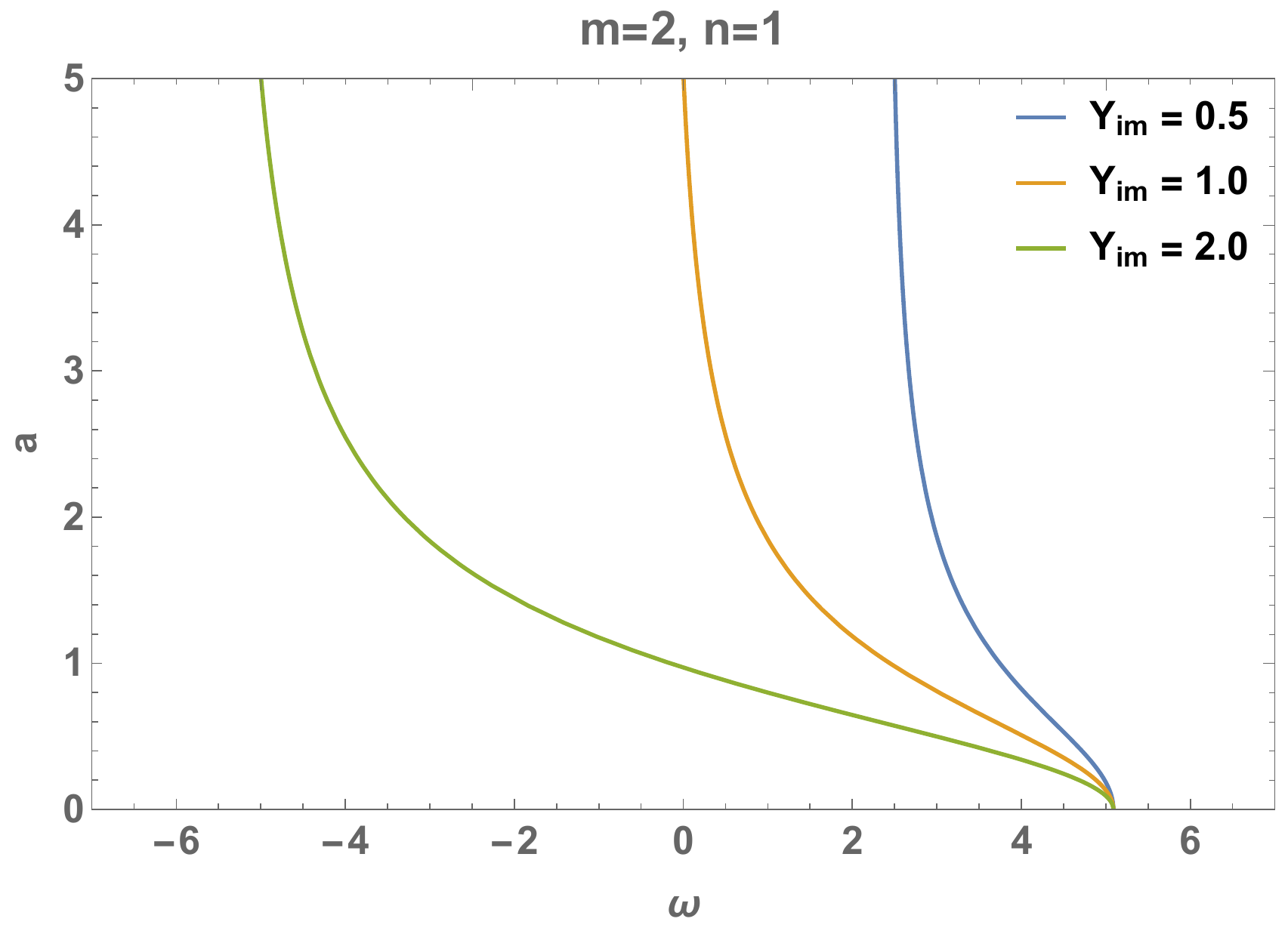}
\includegraphics[width =0.45\columnwidth, height=1.7 in]{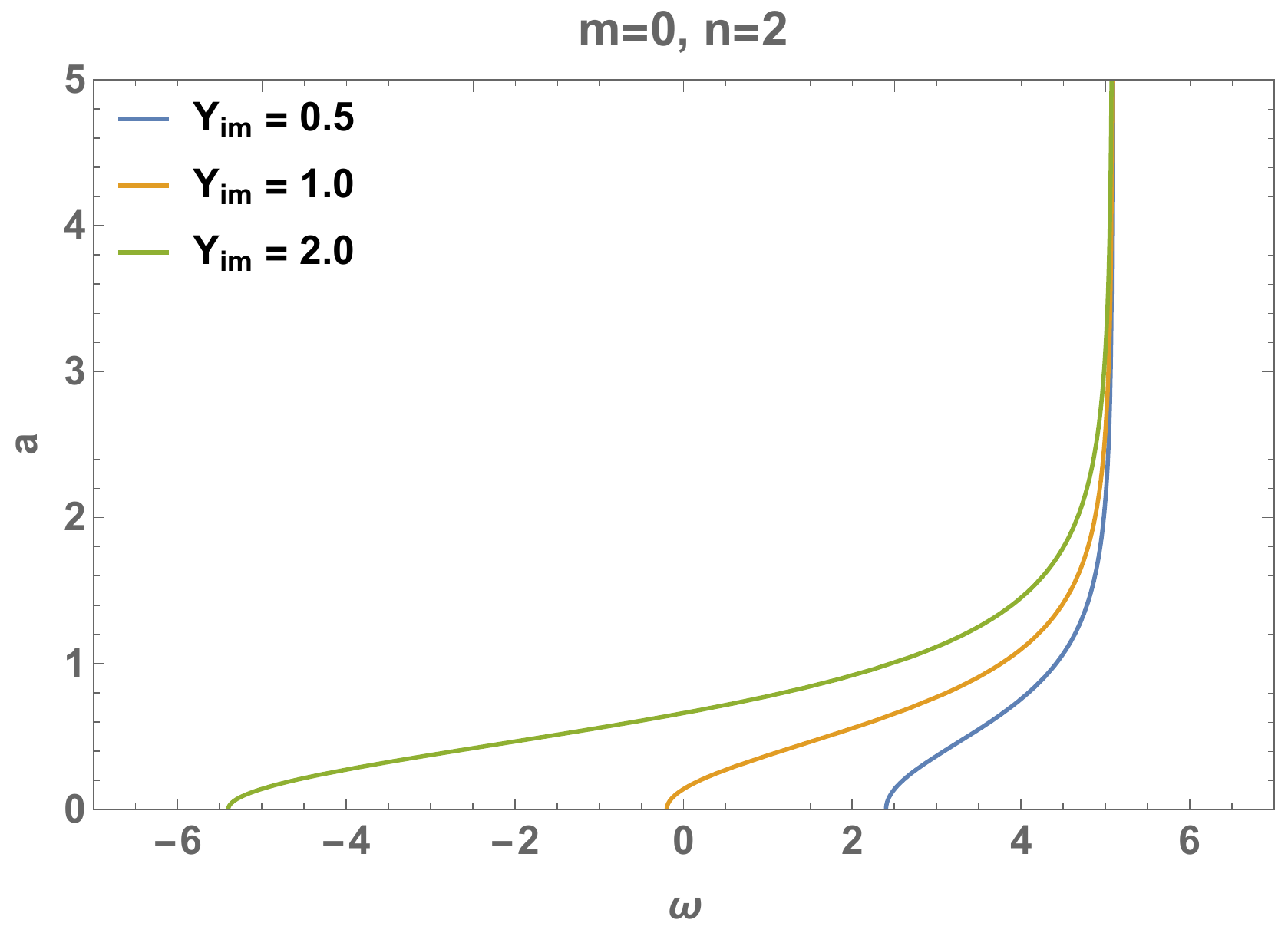}
\includegraphics[width =0.45\columnwidth, height=1.7 in]{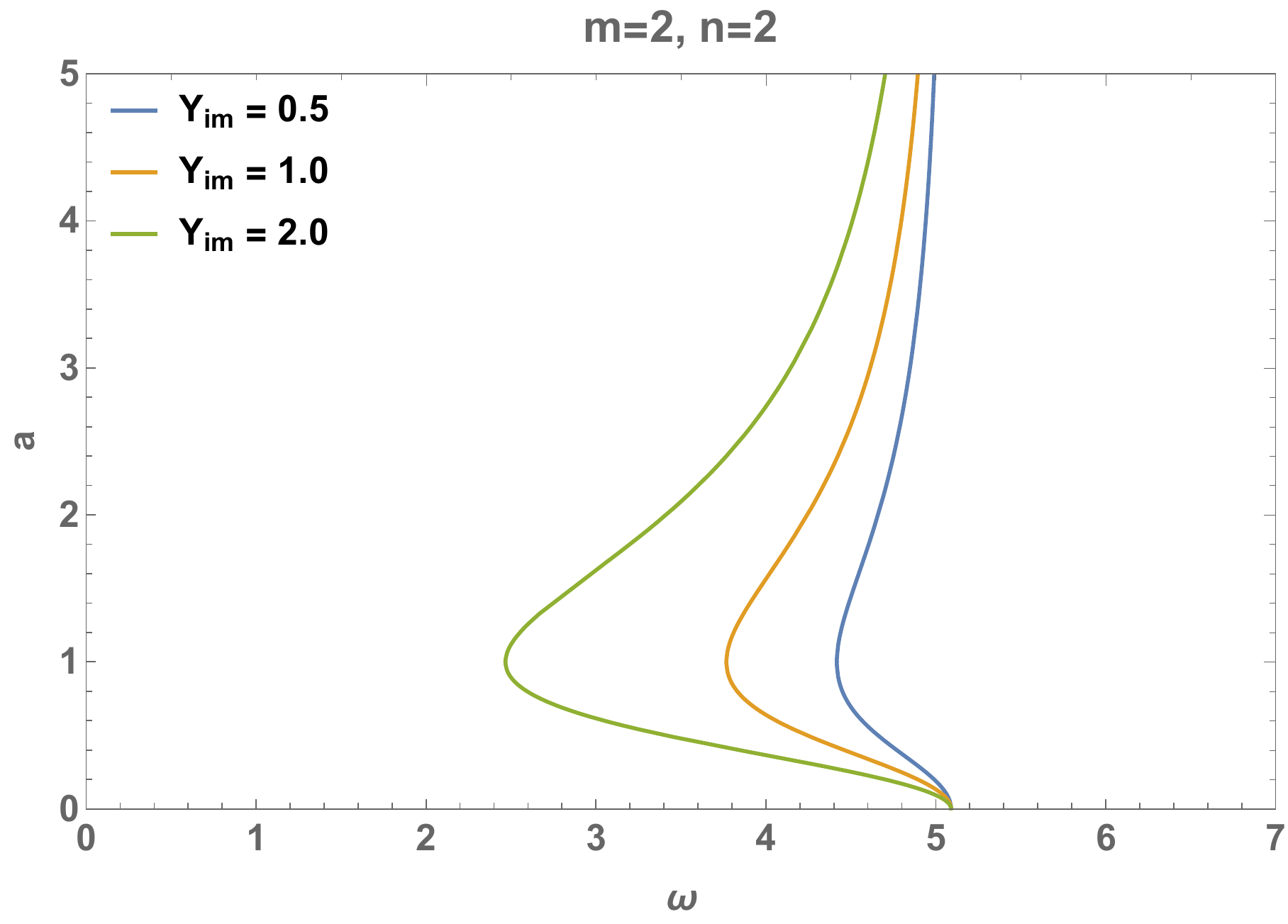}
\end{center}
\caption{(Color online) Loci of the stationary solutions of eq. \ref{eq:fivea} in the upper-half positive a-$\omega$ plane for different values of the pair ($m,n$).} \label{fig:two}
\end{figure}
In the cases $m=0$ but $n=0.5$, $n=1$ and $n=2$, stationary solutions exist only within a limited range of the laser instantaneous frequency $M$ and admit a fixed point at $M=0$, as suggested by eq. (\ref{eq:six}). The nonlinear gain ($-\gamma_r$) also affects the CW solution: As $-\gamma_r$ increases, the maximum value of $M$ decreases correspondingly until no singular solution is permitted. For the Kerr medium i.e. ($m,n$)=($2,0$), and for the case ($m,n$)=($2,1$), the trajectories are different. The loci of the stationary points describe a continuous, increasing function of $M$ with respect to $a$. Therefore these media are more likely to support CW over larger bands than other non-Kerr media. It should be noticed, however, that the saturation order plays an important role in amplitude stabilization by limiting the allowed frequency range for CW solutions. This effect can be noticed in Fig. \ref{fig:one} by the saturation of the instantaneous frequency $M$ for ($m,n$)=($2,1$). For the nonlinearity ($m,n$)=($2,2$) the effect becomes more important, leading to amplitude degeneracy within a specific range of $M$. This behavior is consistent with previous investigations on the stationary solutions of high-order CGL equations, and notably the cubic-quintic CGL equation\cite{a11,FMbieda2020}.\\
Fig. \ref{fig:two} illustrates the stationary solution amplitude on the $\omega - a$ plane. Plot parameters are indicated in the figure. For all the fiber media, stationary solutions exist within a range of $\omega$. In as much the previous paragraph demonstrated a correlation between the CWs and the spectral nonlinear gain ($-\gamma_r$), Fig. \ref{fig:two} also highlights the relationship between the CWs and the nonlinear gain ($\gamma_{im}$) in the CWs propagation constant: The increase in $\gamma_{im}$ widen the range of $\omega$ for which CWs are sustained in the fiber media. Two profiles of the curves can be observed with respect with to the value of $m$ and $n$. Except for the ($2,2$) case, all ($m,n$) pairs have a unique root for each value of $\omega$ within the permitted range. For the ($2,2$) case, two-degenerate fixed points can be observed for every permitted $\omega$, here again in agreement with previous studies \cite{a11,FMbieda2020}.
\section{Modulational-instability analysis}\label{section4}
\begin{figure}[bt]
\begin{center}
\includegraphics[width= 0.45\columnwidth, height= 1.8 in]{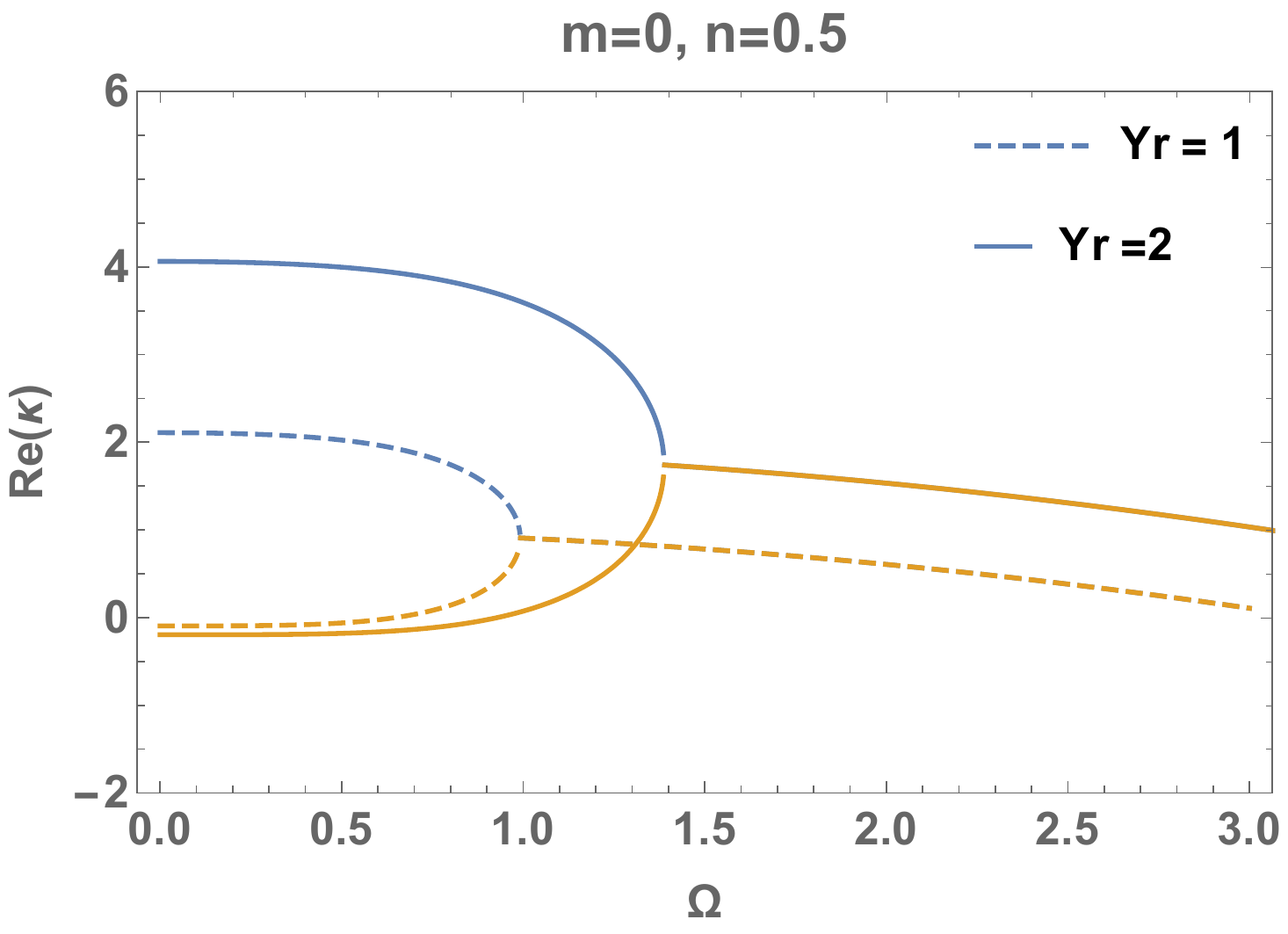}
\includegraphics[width= 0.45\columnwidth, height= 1.8 in]{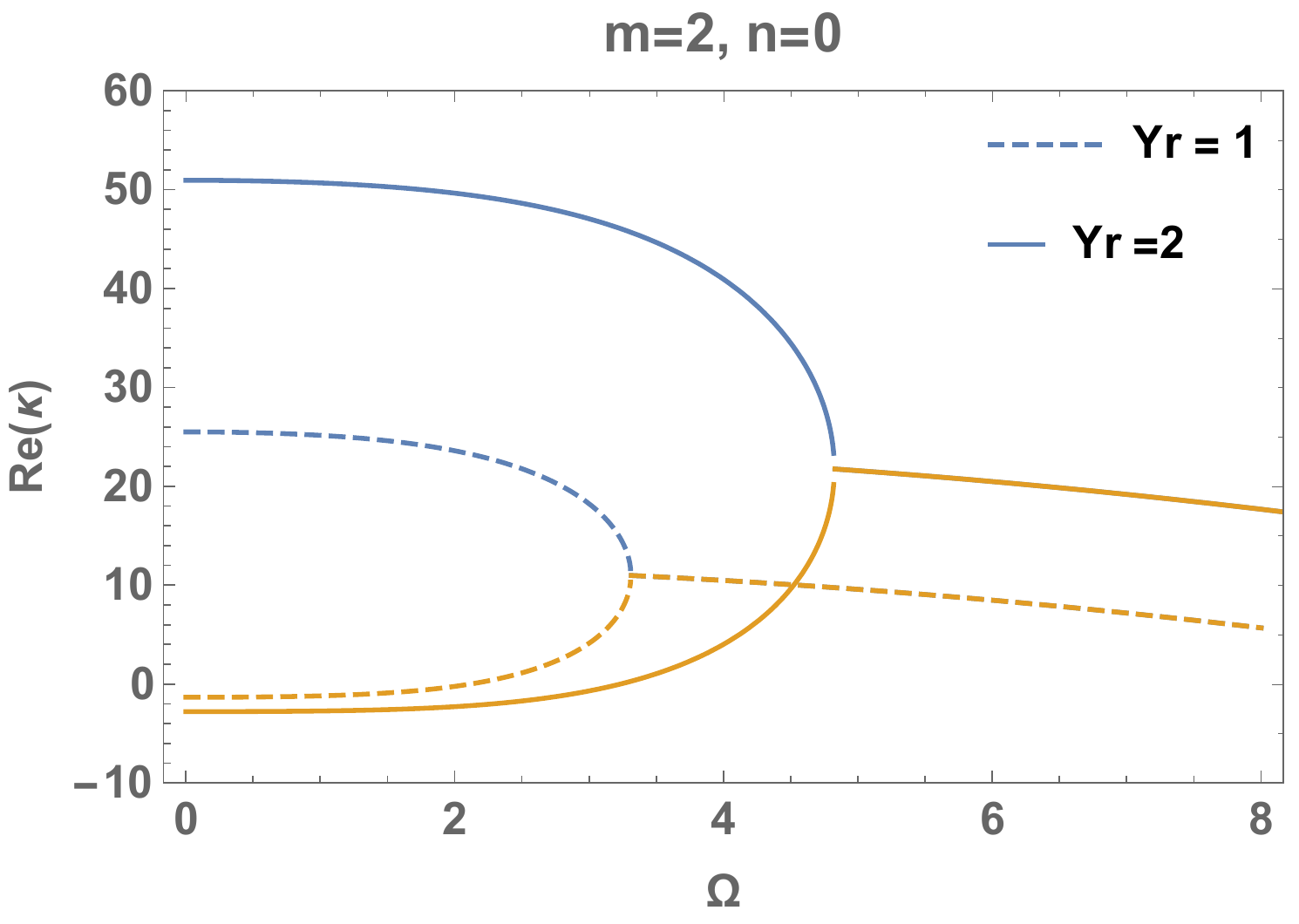}
\includegraphics[width= 0.45\columnwidth, height= 1.8 in]{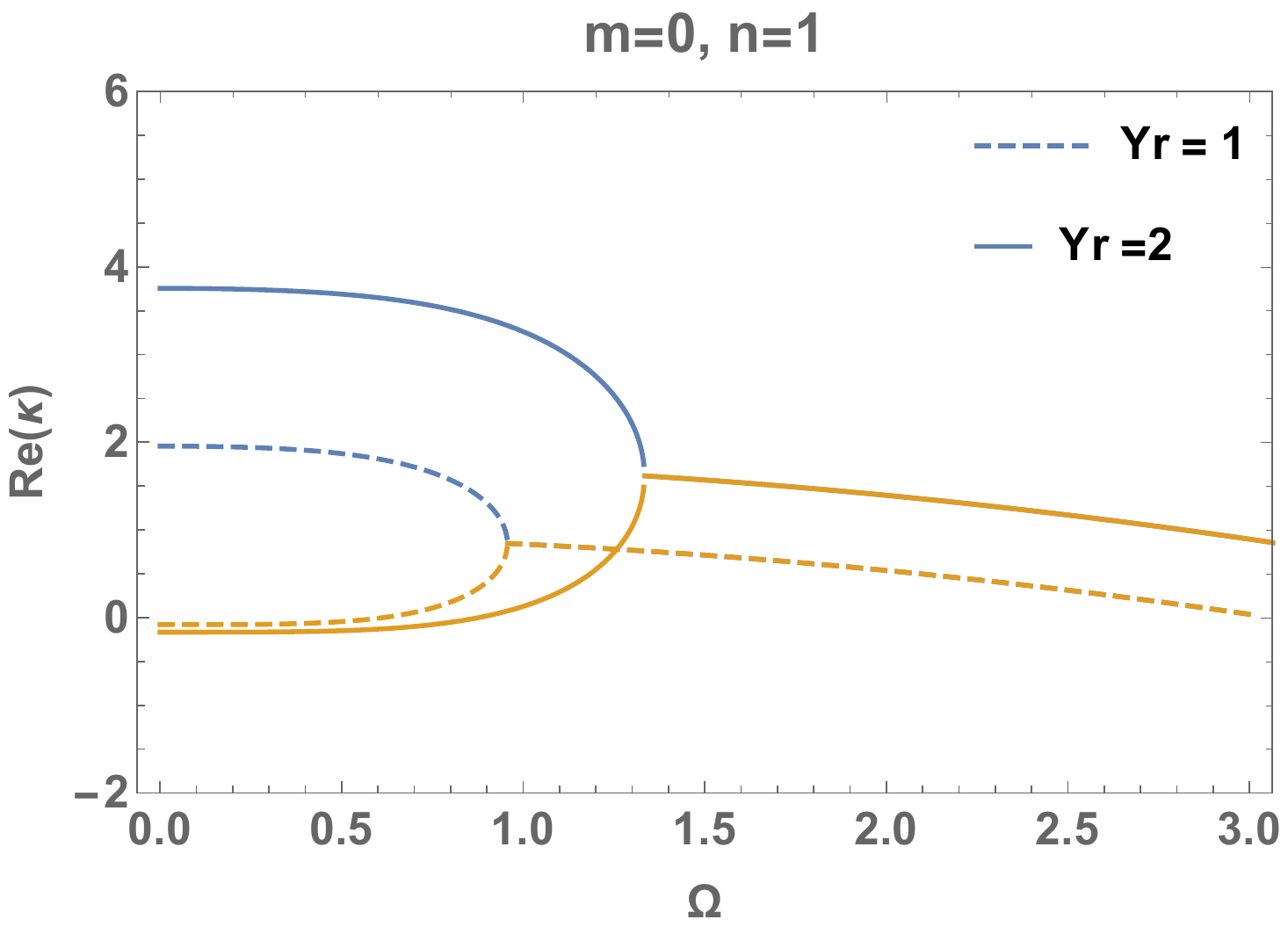}
\includegraphics[width= 0.45\columnwidth, height= 1.8 in]{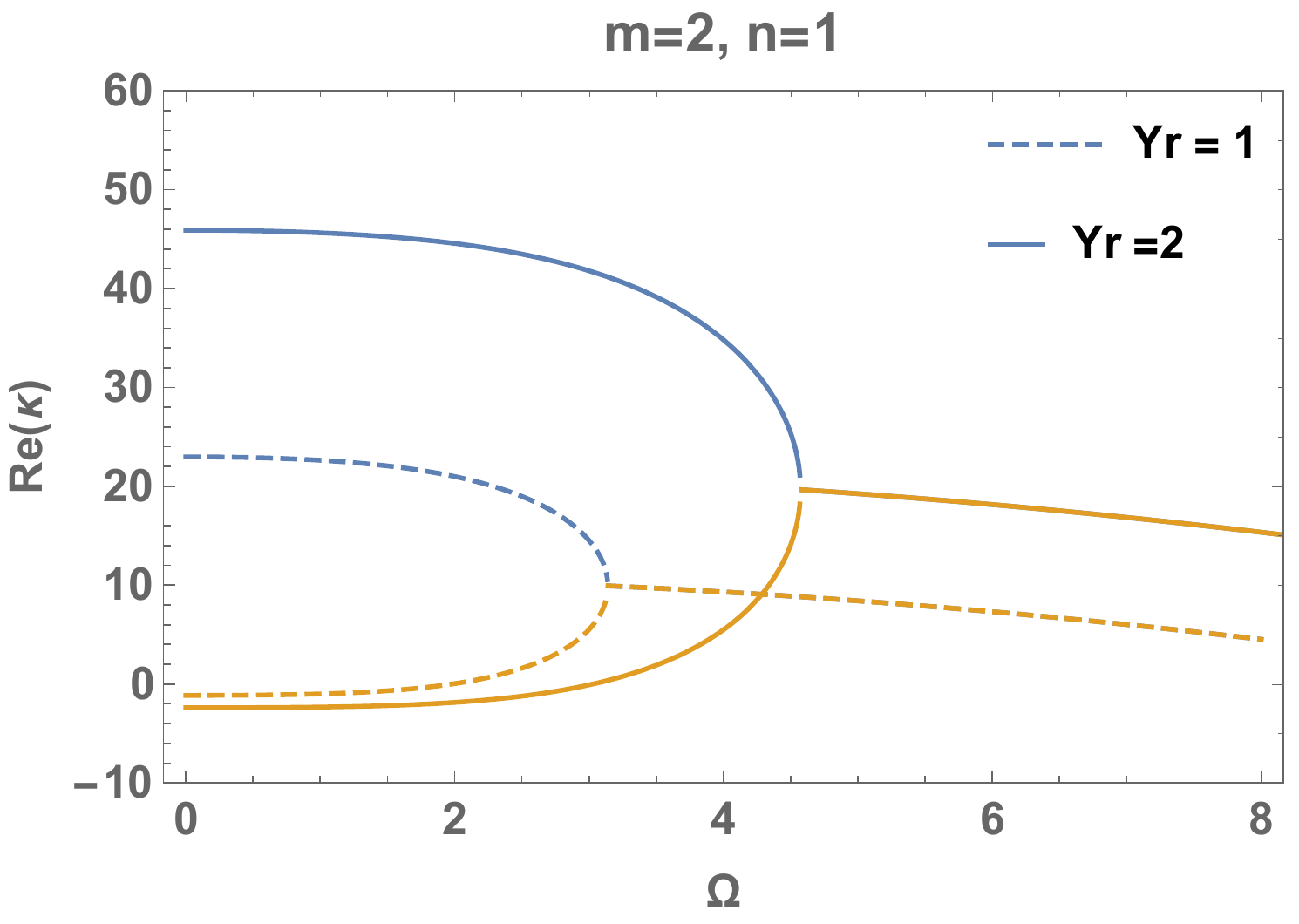} 
\includegraphics[width= 0.45\columnwidth, height= 1.8 in]{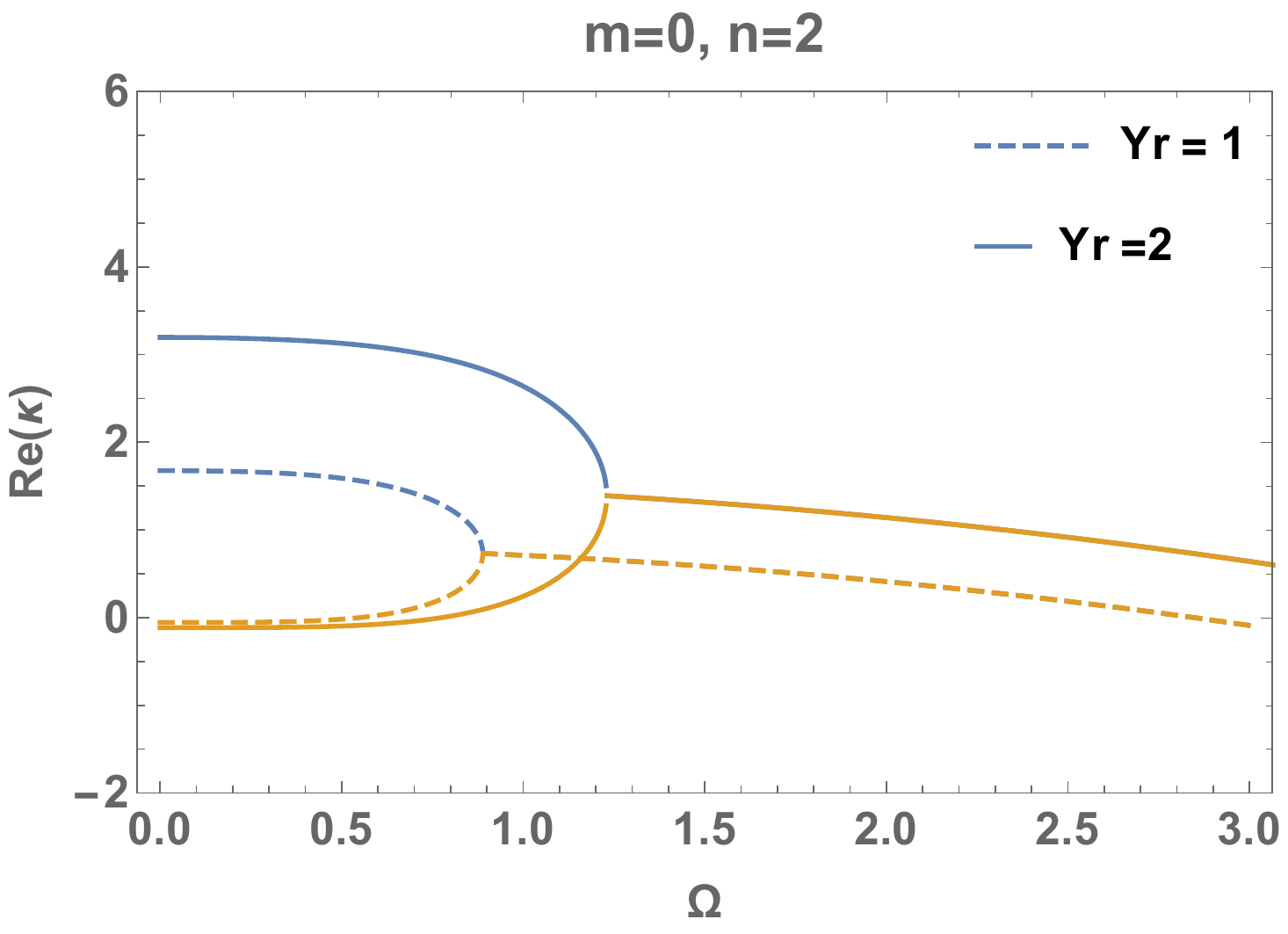} 
\includegraphics[width= 0.45\columnwidth, height= 1.8 in]{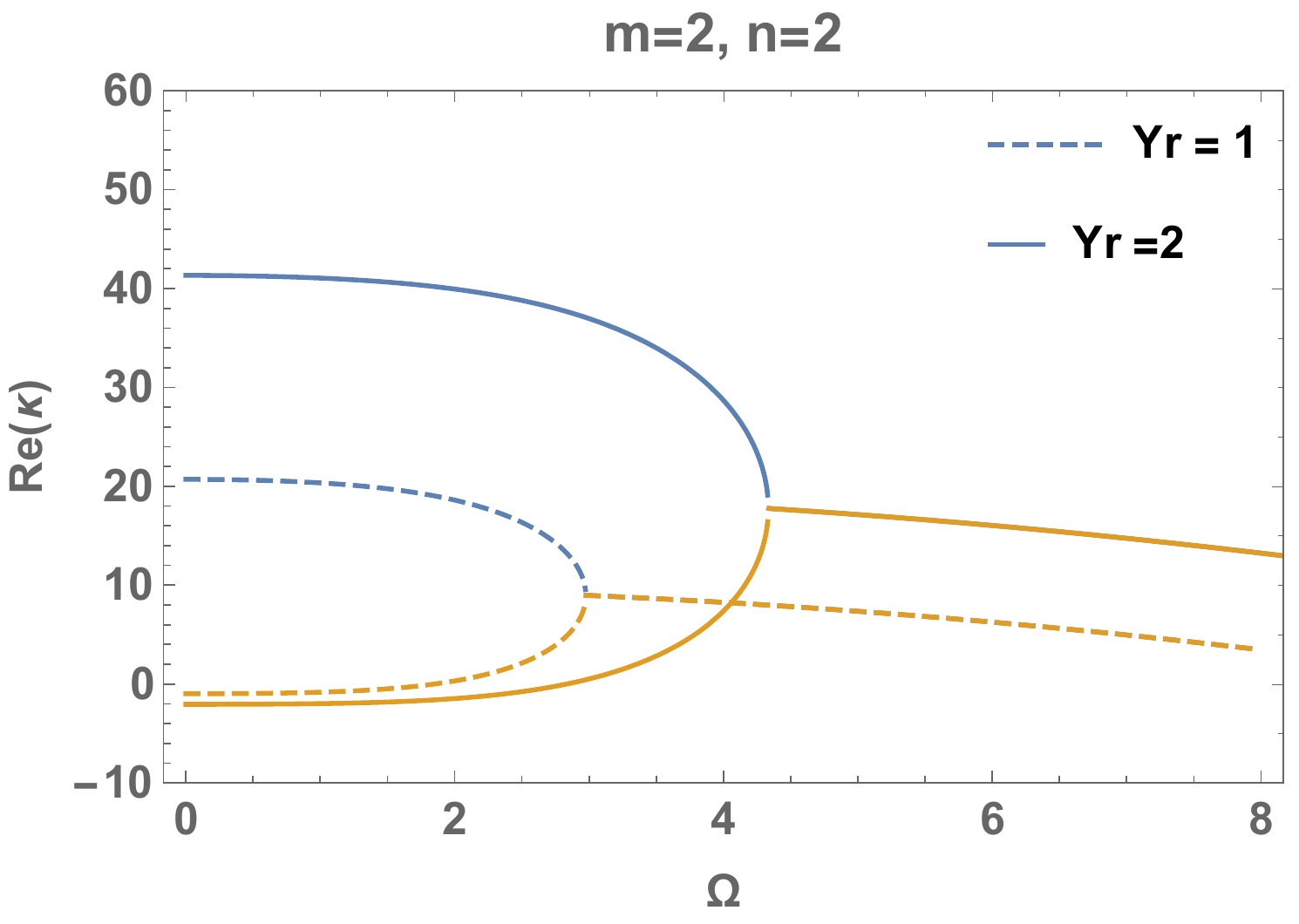} 
\caption{(Color online) Variation of the noise growth rate $Re(\kappa)$ as a function of the noise modulation frequency. Parameter values are: $\theta=0.09$, $\theta=-0.08$, $C=1$, $g-\rho=0.09$}
\label{fig:threea}
\end{center}
\end{figure}

\begin{figure}[bt]
\begin{center}
\includegraphics[width= 0.45\columnwidth, height= 1.8 in]{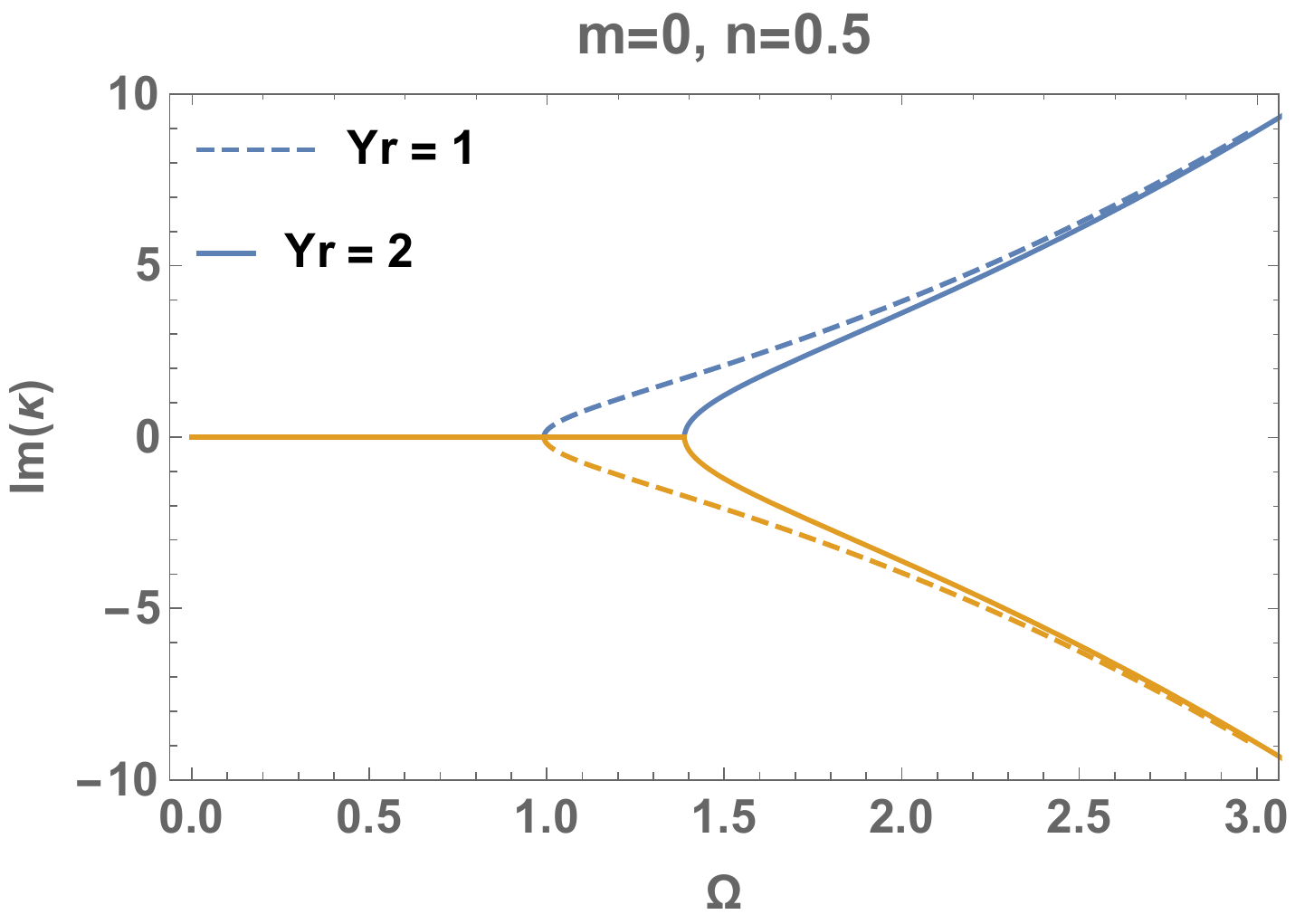}
\includegraphics[width= 0.45\columnwidth, height= 1.8 in]{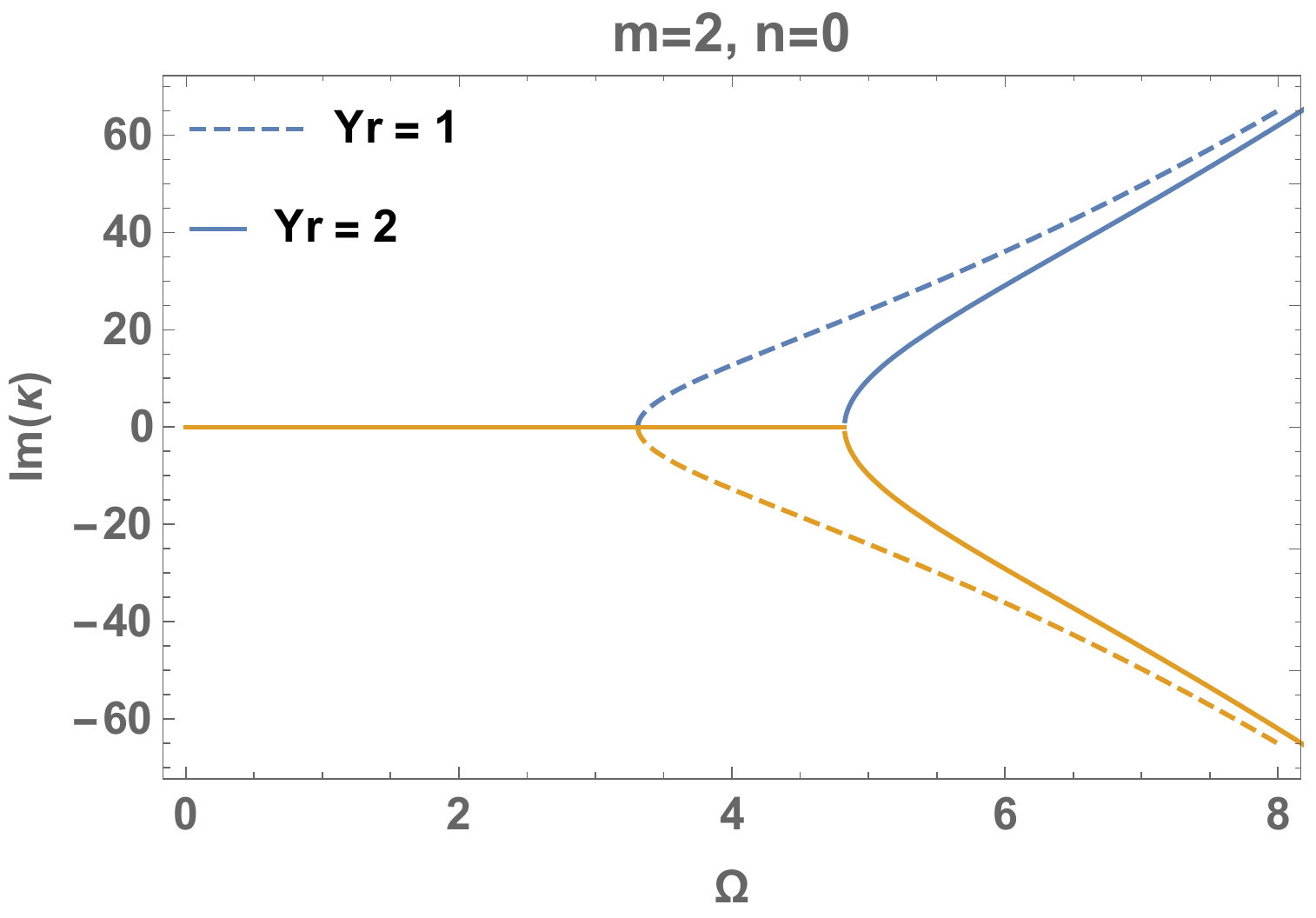}
\includegraphics[width= 0.45\columnwidth, height= 1.8 in]{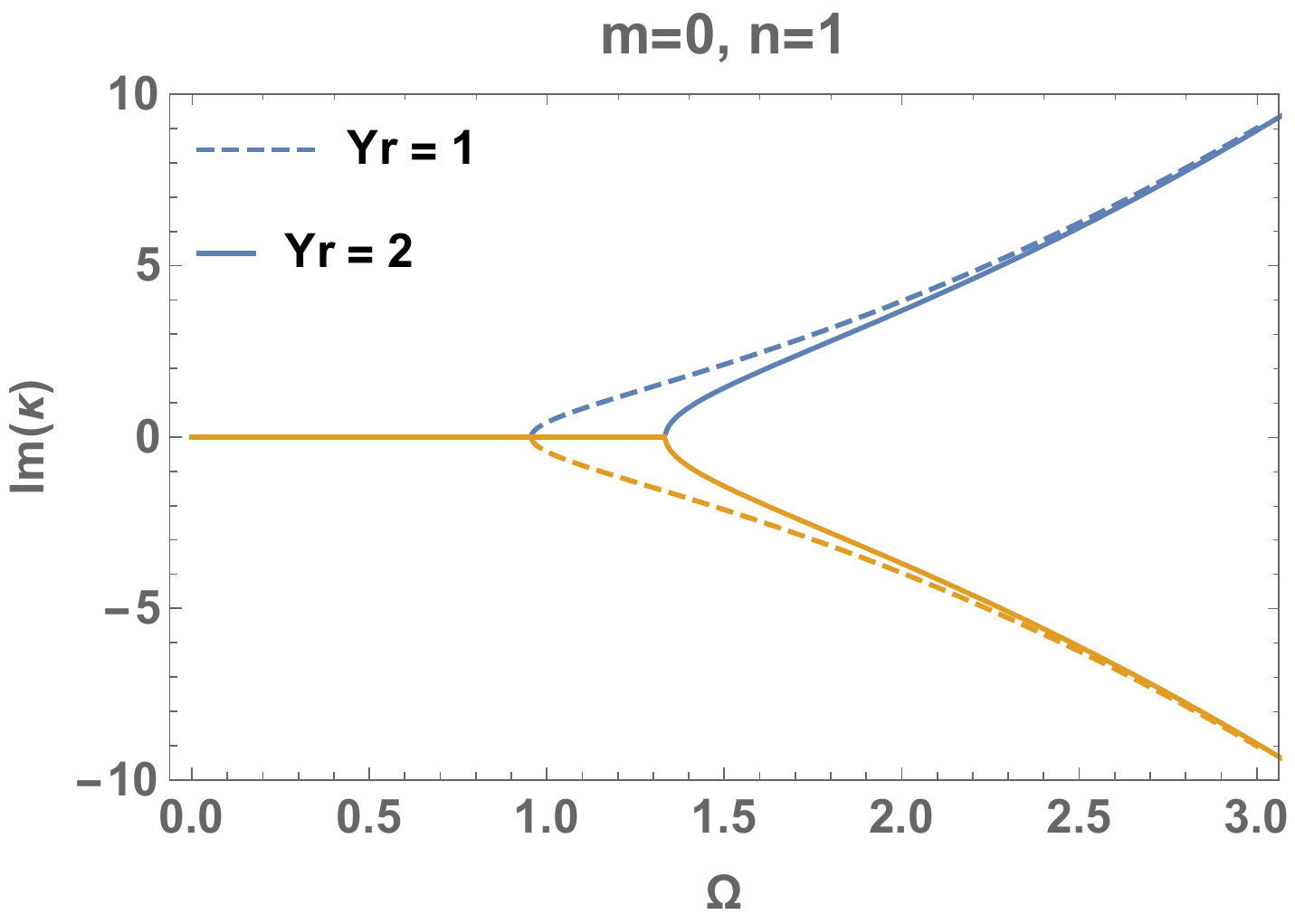}
\includegraphics[width= 0.45\columnwidth, height= 1.8 in]{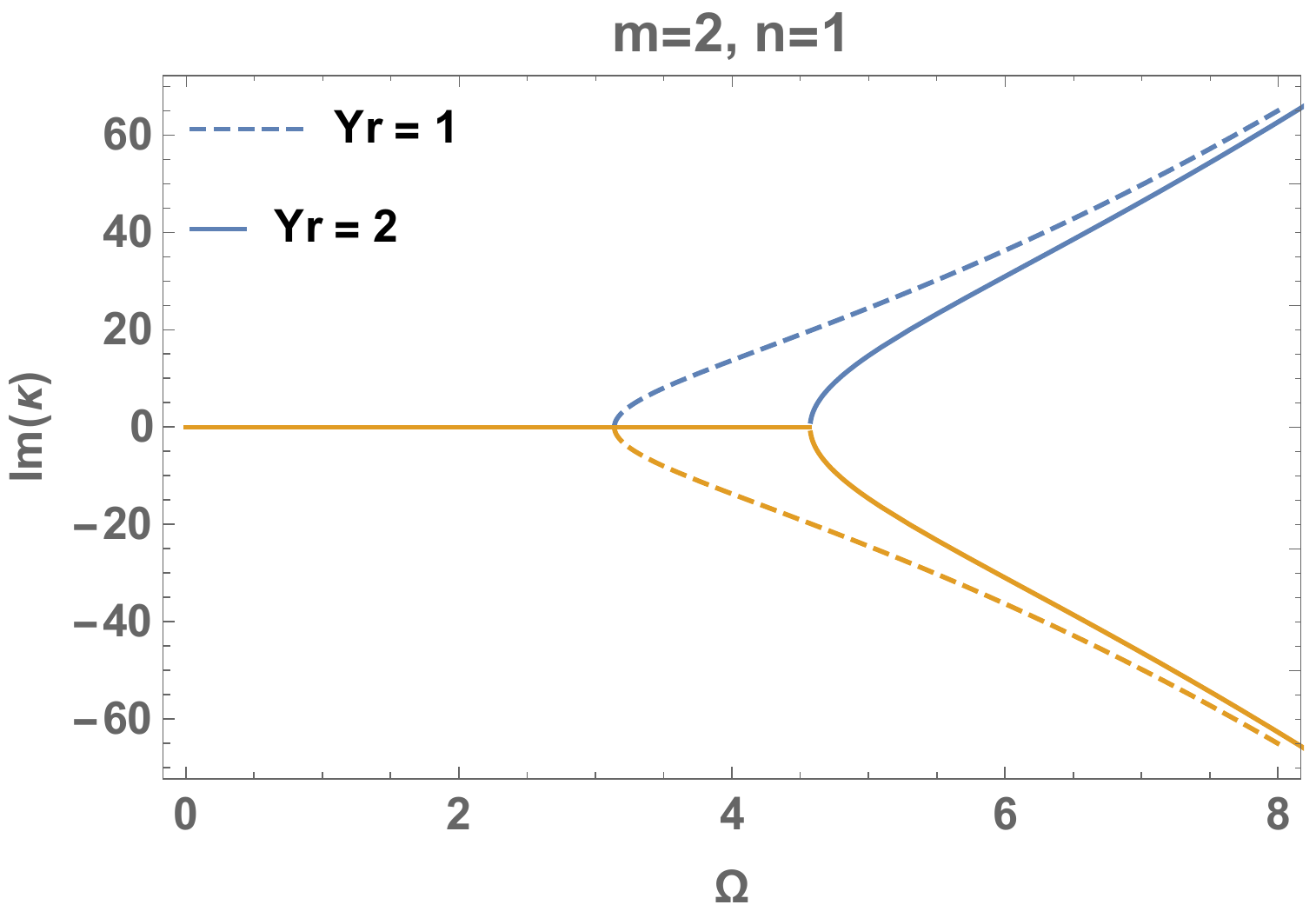} 
\includegraphics[width= 0.45\columnwidth, height= 1.8 in]{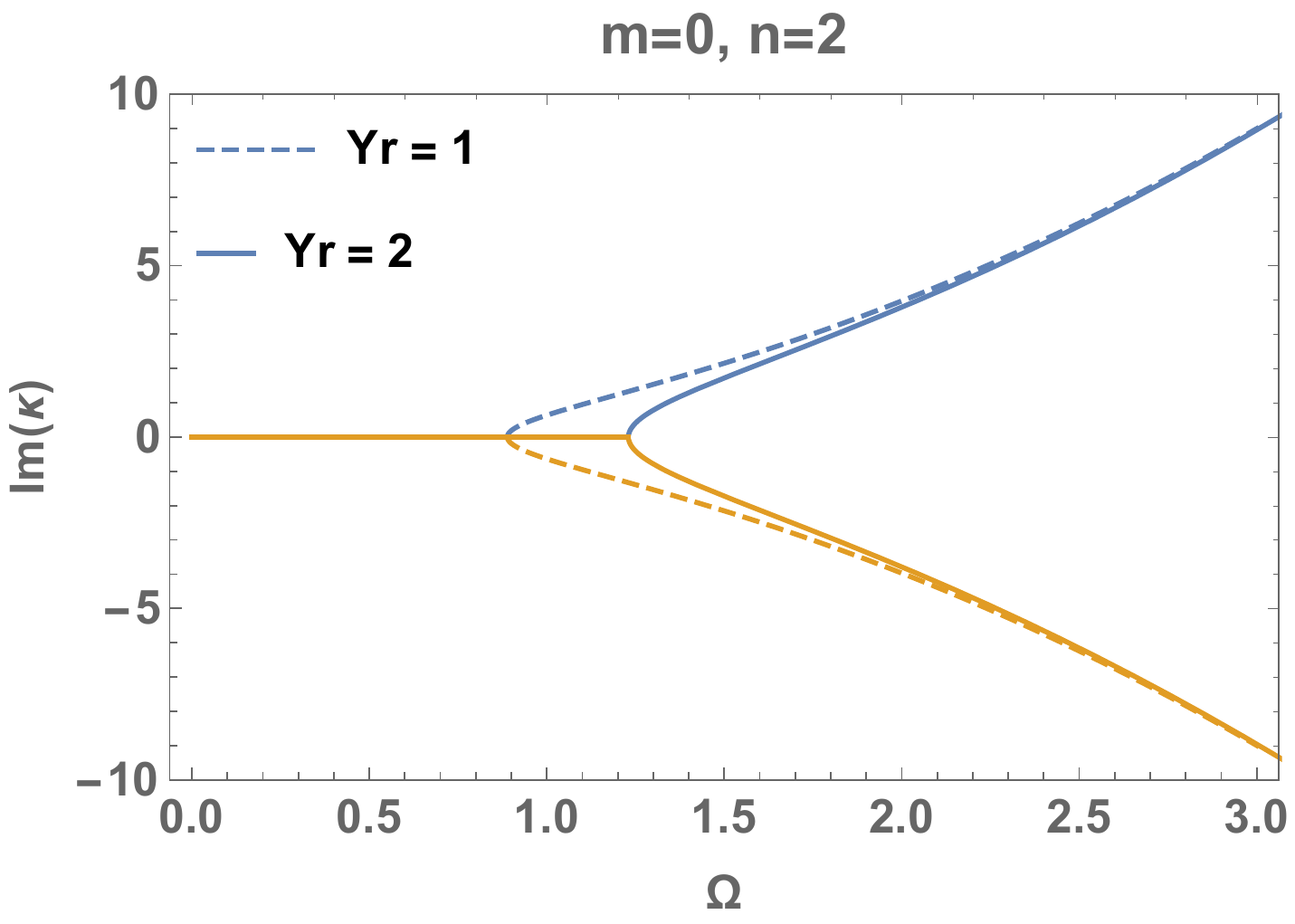} 
\includegraphics[width= 0.45\columnwidth, height= 1.8 in]{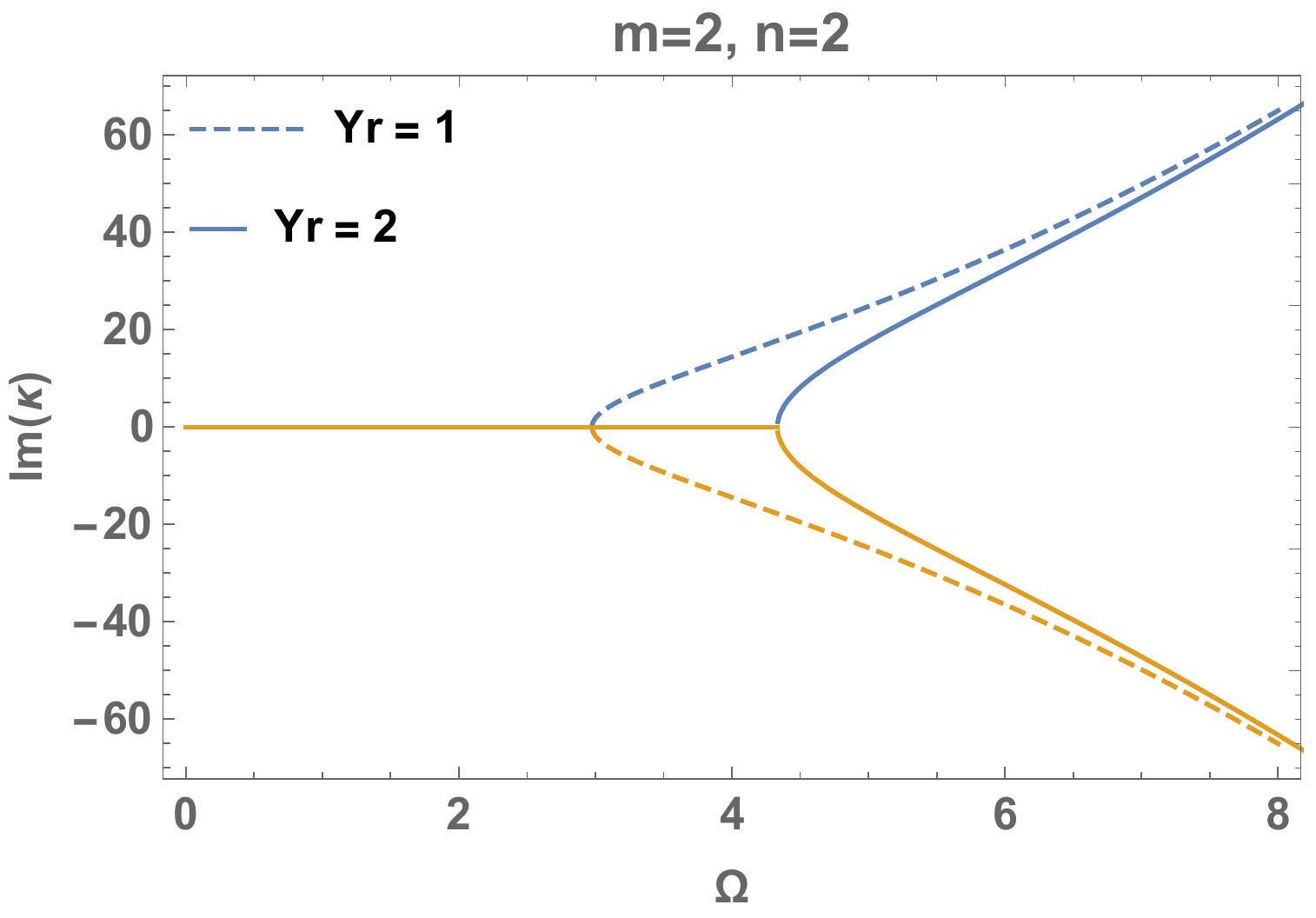} 
\caption{(Color online) Variation of the noise propagation constant (or wave number) $Im(\kappa)$ as a function of the noise modulation frequency $\Omega$. Parameter values are: $\omega=0.09$, $\theta=-0.08$, $C=1$, $g-\rho=0.09$}
\label{fig:threeb}
\end{center}
\end{figure}

The stability of CW solutions to eq. (\ref{eq:one}) can be studied within the context of modulation-instability theory. In this purpose let us consider an harmonic-wave solution of the form: 
\begin{equation}
u(z,t) = a_0 exp\bigg[iM_0 t-i \omega z\bigg],
\end{equation}
 where ($a_0, M_0$) is a stationary fixed-point solution to eqs. (\ref{eq:fivea})-(\ref{eq:fiveb}). The stability of the fixed point ($a_0,M_0$) is determined by the evolution of small perturbation to the CW amplitude. The  perturbed signal can be written as: 
 \begin{equation}
u(z,t) = \bigg[ a_0 exp(iM_0 t) + \epsilon f(t,z)\bigg] exp(-i \omega z), \label{eq:seven}
\end{equation}
 where $\epsilon$ is the perturbation strength (assumed small compared to $a_0$) and $f(t,z)$ is the noise function. By substituting eq. (\ref{eq:seven}) into eq. (\ref{eq:one}) and retaining only linear terms in $\epsilon$, eq. (\ref{eq:one}) can be rewritten:
\begin{eqnarray}
 f_z &=& (C+iD)f_{tt}+ \left[ (g-\rho) + i(\omega+\theta)\right]f \nonumber \\ 
 &+& \frac{(\gamma_r + i \gamma_{im})a^{m}_0}{(1+\Gamma a^2_0)^n} f \nonumber \\
 &+& \frac{(\gamma_r + i \gamma_{im})a^{m+1}_0}{2} \left(\frac{m}{a_0(1+\Gamma a^2_0)^n}- \frac{2 n \Gamma a_0}{(1+\Gamma a^2_0)^{n+1}}\right) \nonumber \\
 &\times& (f e^{-i M_0 t} + f^* e^{i M_0 t}),\label{eq:eight}
\end{eqnarray}
 where the asterisk denotes the complex conjugate. A similar equation can be derived for the complex conjugate of $f$, i.e. $f^*$, in such a way that a set of coupled linear equations for $f$ and $f^*$ is obtained. Consider a noise function of the form: 
 \begin{equation}
 [f(t,z), f^*(t,z)] = [A_1(\kappa,\Omega),A_2(\kappa,\Omega)] exp(-i \Omega t + \kappa z), \label{eq:nine}
 \end{equation}
where $\Omega$ is the modulation frequency and $\kappa$ is the spatial amplification coefficient of noise. By substituting eq. (\ref{eq:nine}) into eq. (\ref{eq:eight}) and proceeding similarly with the corresponding complex conjugate equation, in the steady state $M_0=0$ the eigenvalue equations for the noise amplitudes can be rewritten in matrix form as:
\begin{equation}
\left(\begin{array}{cc}
i \kappa +Q & P \\ 
P^* & Q^*-i \kappa
\end{array}\right)
\left(\begin{array}{c}
A_1  \\ 
A_2
\end{array}\right) =
\left(\begin{array}{c}
0  \\ 
0
\end{array}\right), \label{eq:ten}
\end{equation}
where:
\begin{eqnarray}
P &=& \frac{(\gamma_{im} - i \gamma_{r})}{2} \bigg [\frac{m}{a_0(1+\Gamma a^2_0)^n}- \frac{2 n \Gamma a_0}{a_0(1+\Gamma a^2_0)^{n+1}}\bigg] a^{m+1}_0, \nonumber \\
Q &=& (\omega + \theta) - i (g-\rho) + (iC-D)\Omega^2 + P \label{eq:eleven}
\end{eqnarray}
Non-trivial solutions to eq. (\ref{eq:nine}) requires the determinant of the 2$\times$ 2 square matrix in eq. (\ref{eq:ten}) to be zero. This condition leads to the secular equation:
\begin{equation}
\kappa^2-2 Im(Q) \kappa + |Q|^2-|P|^2 = 0,
\end{equation}
that admits two roots namely: 
\begin{equation}
\kappa = - Im(Q) \pm \sqrt{|P|^2-Re(Q)}, \label{eq:twelve}
\end{equation}
where $Re(Q)$ and $Im(Q)$ refer to the real and imaginary parts of $Q$, respectively. Since $\kappa$ is a complex function, its real part $Re(\kappa)$ is equivalent to the noise spatial growth rate whereas its imaginary part $Im(\kappa)$ corresponds to the noise propagation constant (or wave number). Fig. \ref{fig:threea} and Fig. \ref{fig:threeb} sketch the evolution of $Re(\kappa)$ and $Im(\kappa)$ as function of the modulation frequency, for various nonlinear laws ($m,n$) indicated in the graphs. Curves were obtained for $\theta=0.09$, $\theta=-0.08$, $C=1$ and $g-\rho=0.09$.

It is quite remarkable that for all the values of the pair ($m,n$) considered, the growth rate $Re(\kappa)$ increases as the modulation frequency decreases until a bifurcation point $\Omega_{th}$, where $Re(\kappa)$ degenerates into two branches: the increasing branch in the graphs corresponds to the positive root, and the decreasing branch to the negative root in formula (\ref{eq:twelve}). For the parameters considered in Fig. \ref{fig:threea} and Fig. \ref{fig:threeb}, the value of $\Omega_{th}$ can be tuned by the nonlinear gain ($-\gamma_r$). Moreover, $\Omega_{th}$ does not vary much with respect to $n$ but increases with $m$. Also, the growth rate $Re(\kappa)$ and the propagation constant $Im(\kappa)$ are increased significantly at larger $m$. \\
The bifurcation of the noise growth rate and propagation constant can be associated with the generation of two CW solutions: One of these solutions has a negative value of $Re(\kappa)$ at low $\Omega$ and hence is expected to be stable, whereas the other solution has a positive value of $Re(\kappa)$ causing and instability of CW. As emphasized in previous works \cite{a6,FMbieda2020,kam},  pulse and multi-pulse like structures are more likely to form by Kerr or higher-order nonlinear effects through the process of modulational instability. Indeed the unstable CW signal when $Re(\kappa)$ $>0$, can break into multi-pulse structures as a result of competing dispersion and  nonlinearities within the optical medium. Our findings are consistent with previous modulational-instability investigations namely in the context of Kerr nonlinearity i.e. ($2,0$), and for the physical context ($2,1$) or ($2,2$) this later case sharing most of the peculiar features of the cubic-quintic CGL equation \cite{a6,a11,FMbieda2020,kam}. 

\section{Pulse and multi-pulse solutions}\label{section5}
\begin{figure}[h]
\begin{center}
\includegraphics[width= 0.45\columnwidth, height= 1.8 in]{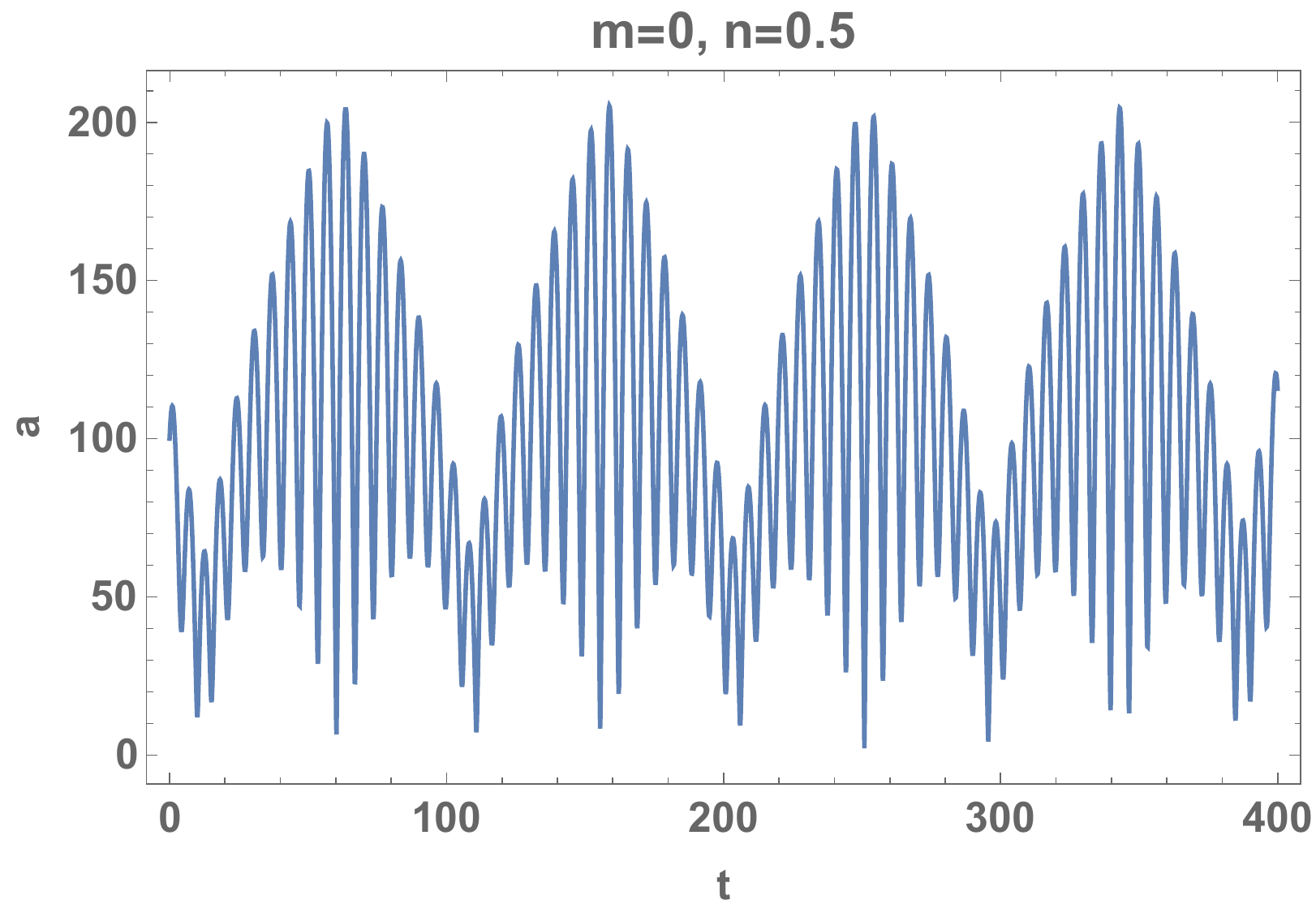}
\includegraphics[width= 0.45\columnwidth, height= 1.8 in]{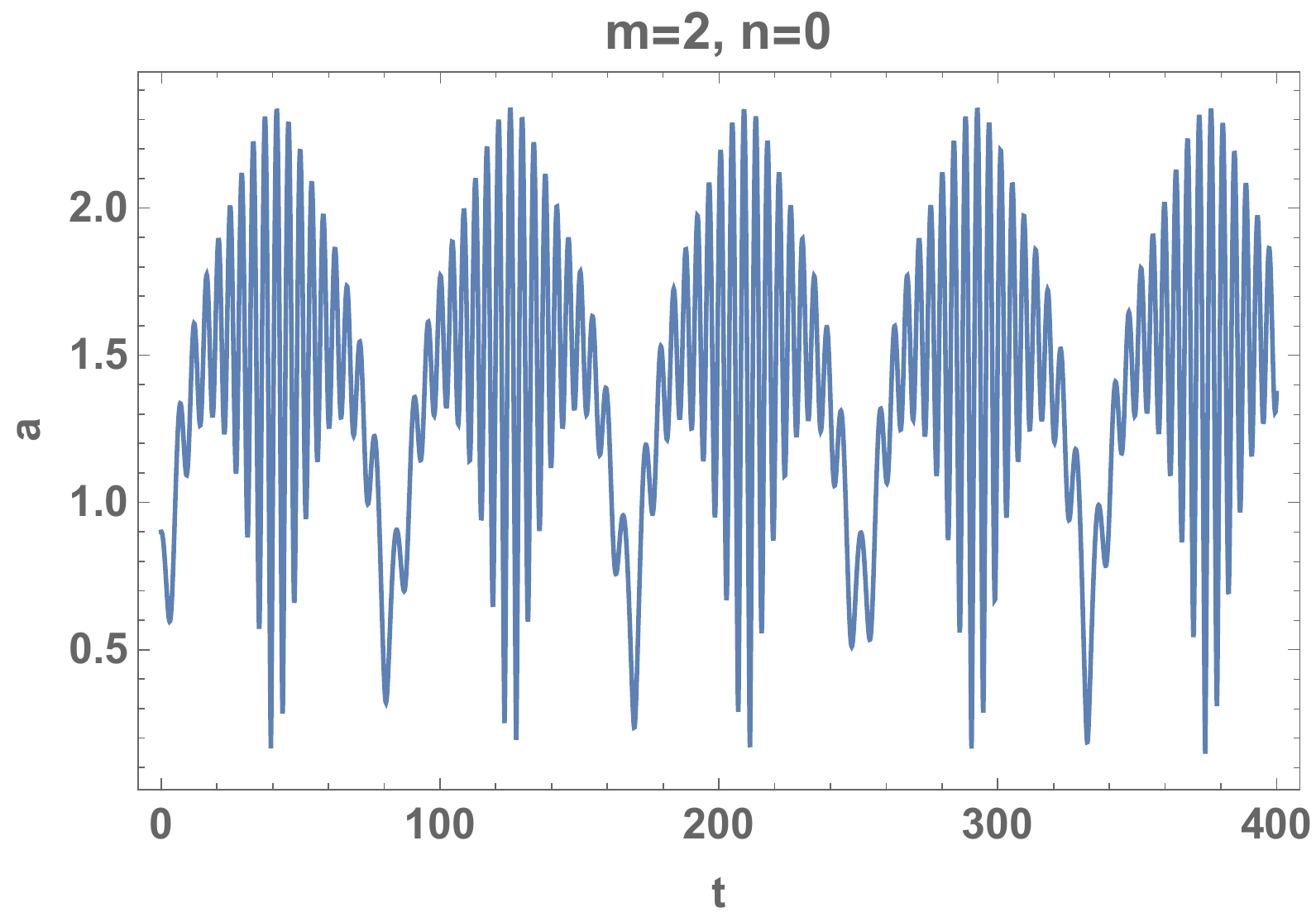}
\includegraphics[width= 0.45\columnwidth, height= 1.8 in]{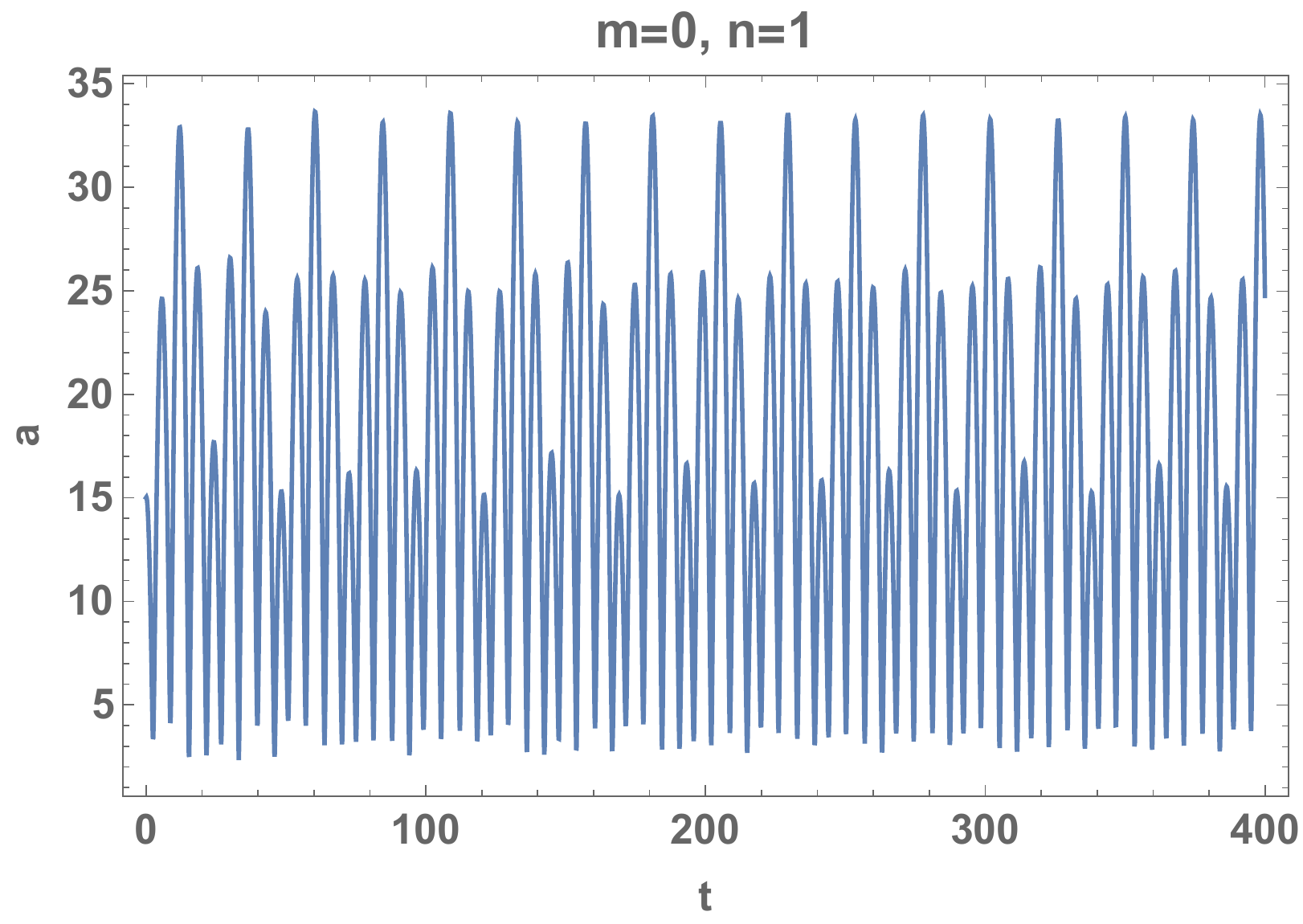}
\includegraphics[width= 0.45\columnwidth, height= 1.8 in]{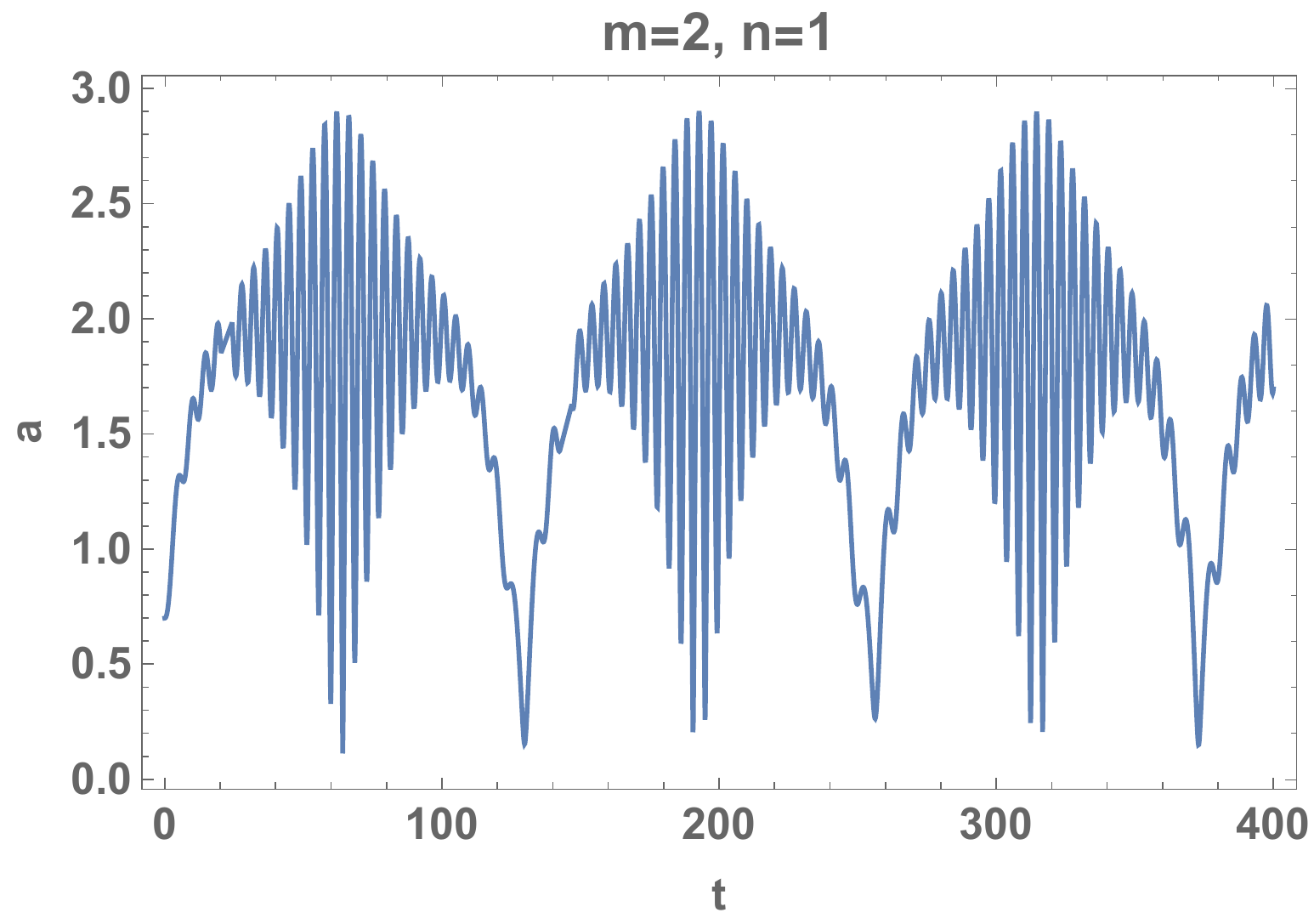} 
\includegraphics[width= 0.45\columnwidth, height= 1.8 in]{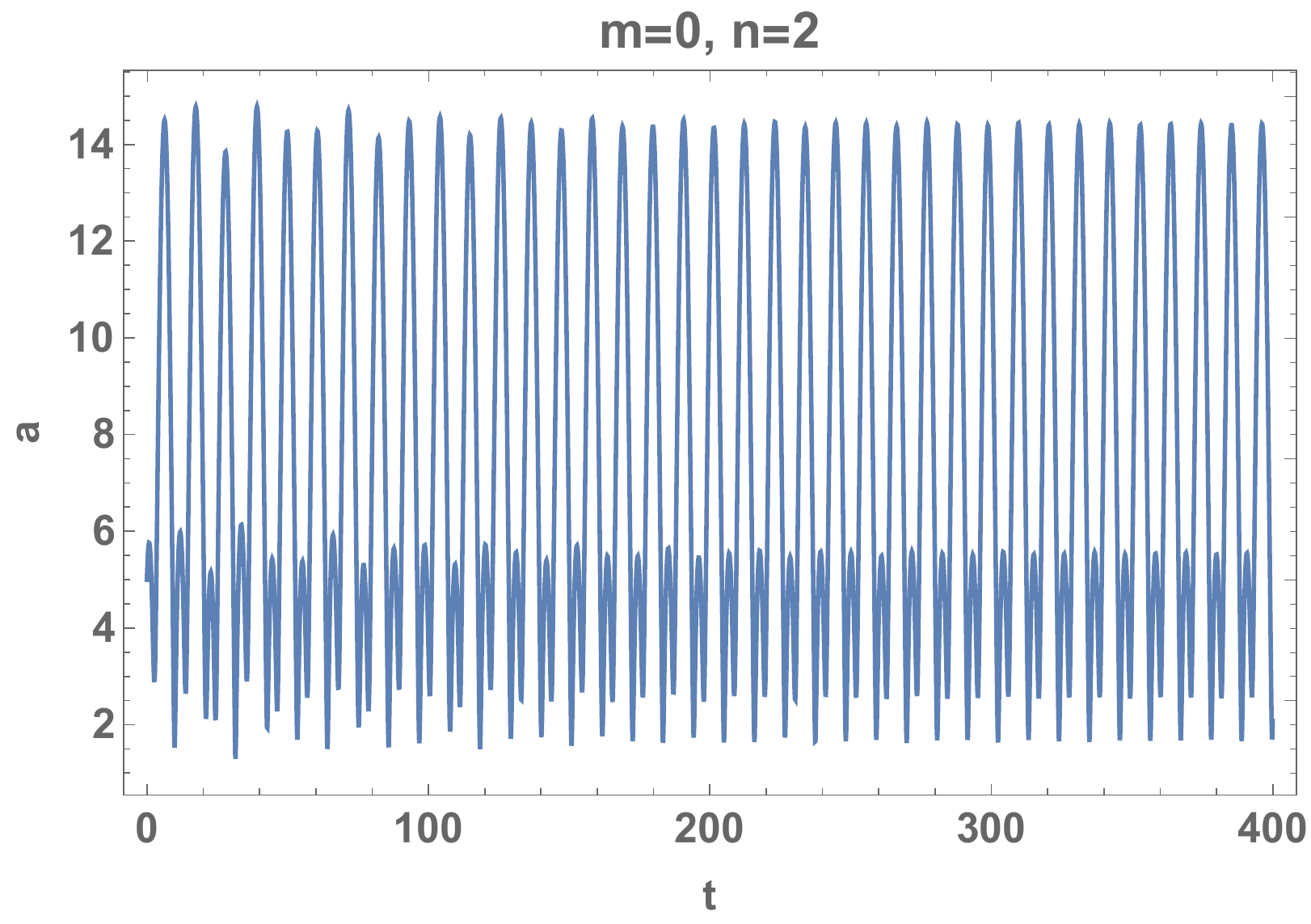} 
\includegraphics[width= 0.45\columnwidth, height= 1.8 in]{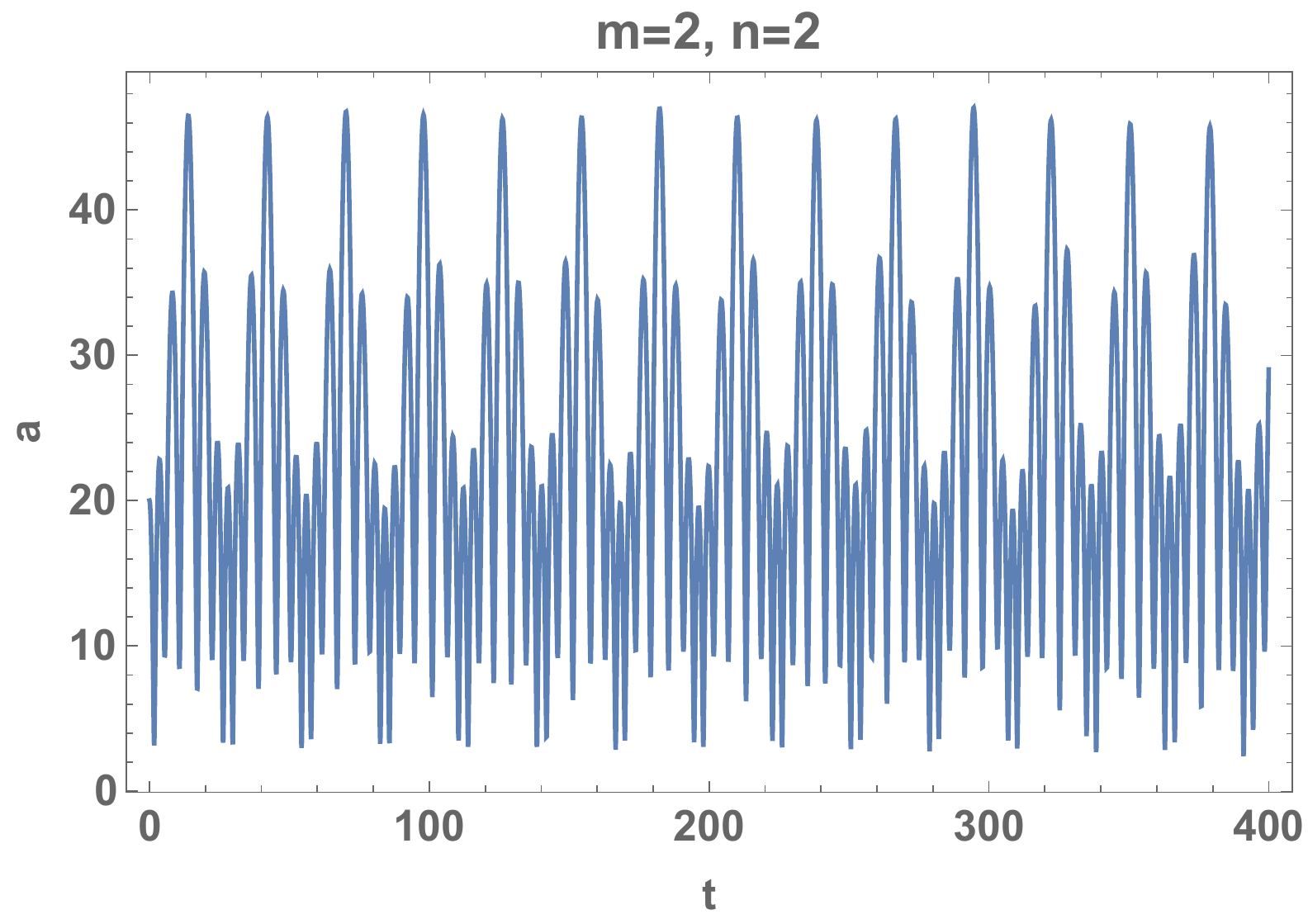} 
\caption{(Color online) Multi-pulse patterns emerging from time evolution of the laser amplitude $a(t)$, obtained from numerical simulations of eqs. (\ref{eq:foura})-(\ref{eq:fourd}). Parameter values are: $\omega=0.9$, $\theta=-0.7$ , $C=5$, $D=1$, $\gamma_r=0.7$, $\gamma_{im}=-0.14$, and $\Gamma=0.1$ for ($2,0$) and ($2,1$) media, and  $\gamma_{r}$=0.7, $\Gamma=0.02$ for ($2,2$). These parameters were varied near the values  $\omega=0.14$, $\theta=0.1$ , $C=1$, $D=1$, $\gamma_r=5$, $\gamma_{im}=-0.14$, and $\Gamma=0.04$ for ($2,0$) and ($2,1$) for ($0,0.5$), ($0,1$) and ($0,2$) pairs.}
\label{fig:four}
\end{center}
\end{figure}

\begin{figure}[h]
\begin{center}
\includegraphics[width= 0.45\columnwidth, height= 1.8 in]{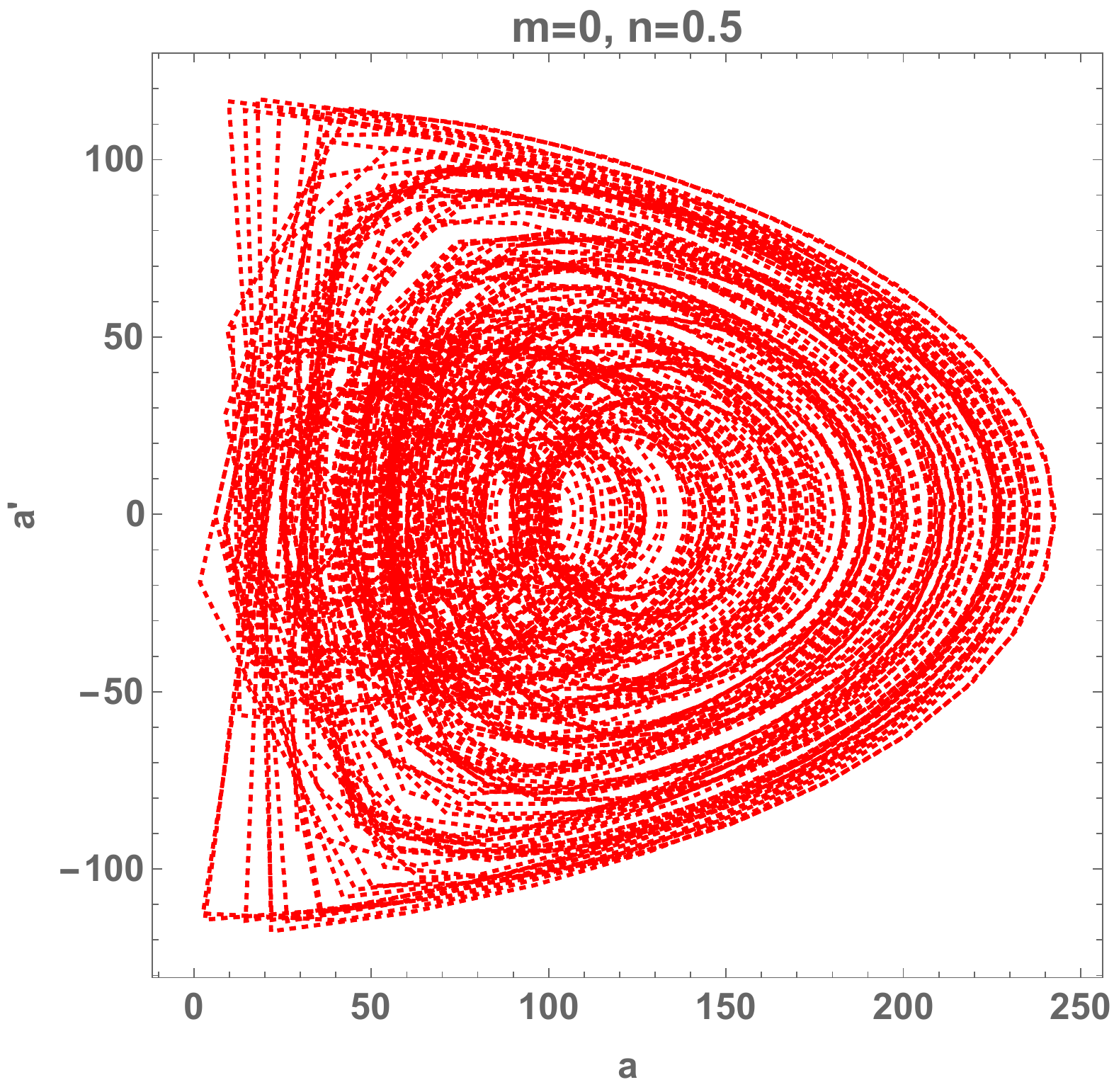}
\includegraphics[width= 0.45\columnwidth, height= 1.8 in]{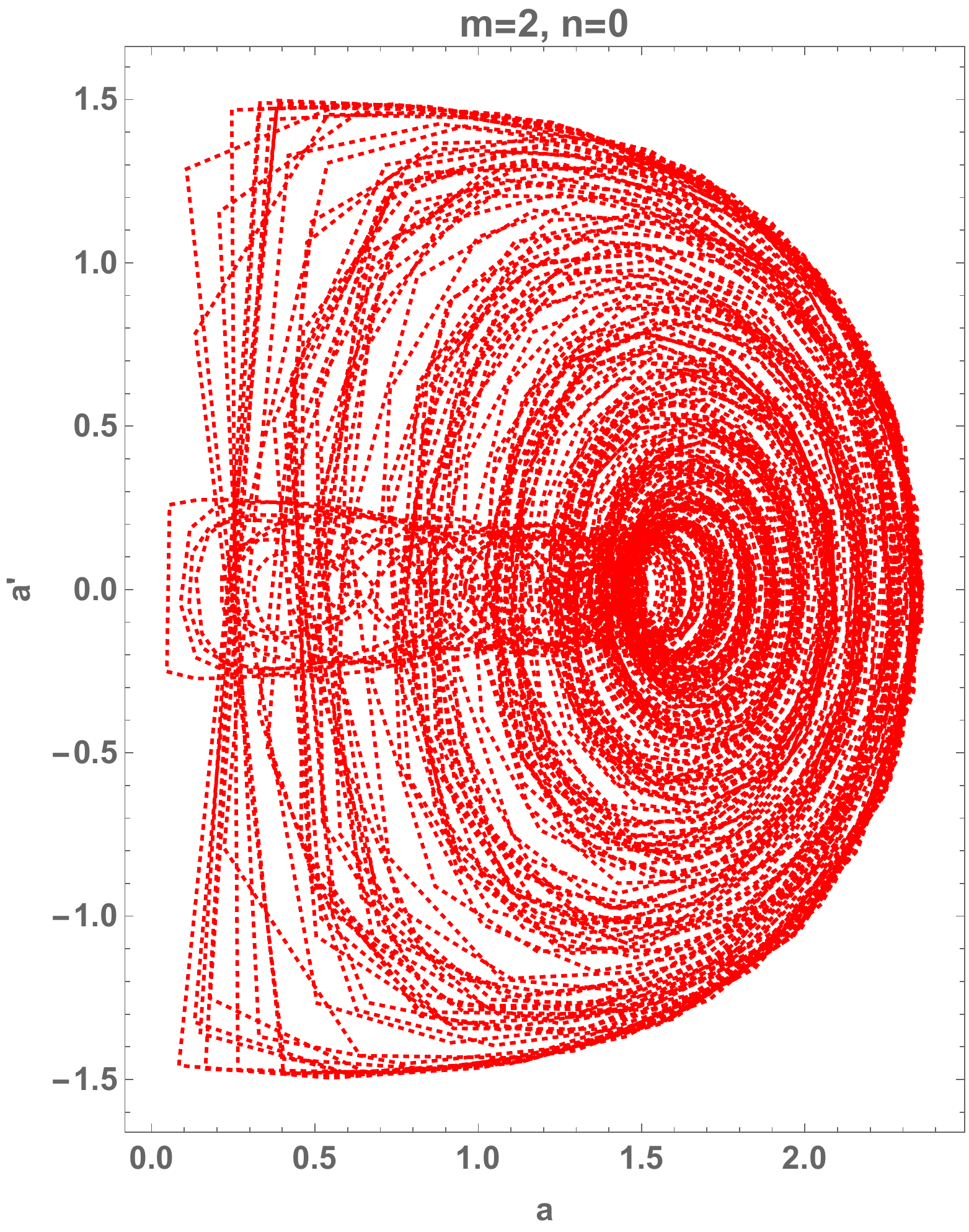}
\includegraphics[width= 0.45\columnwidth, height= 1.8 in]{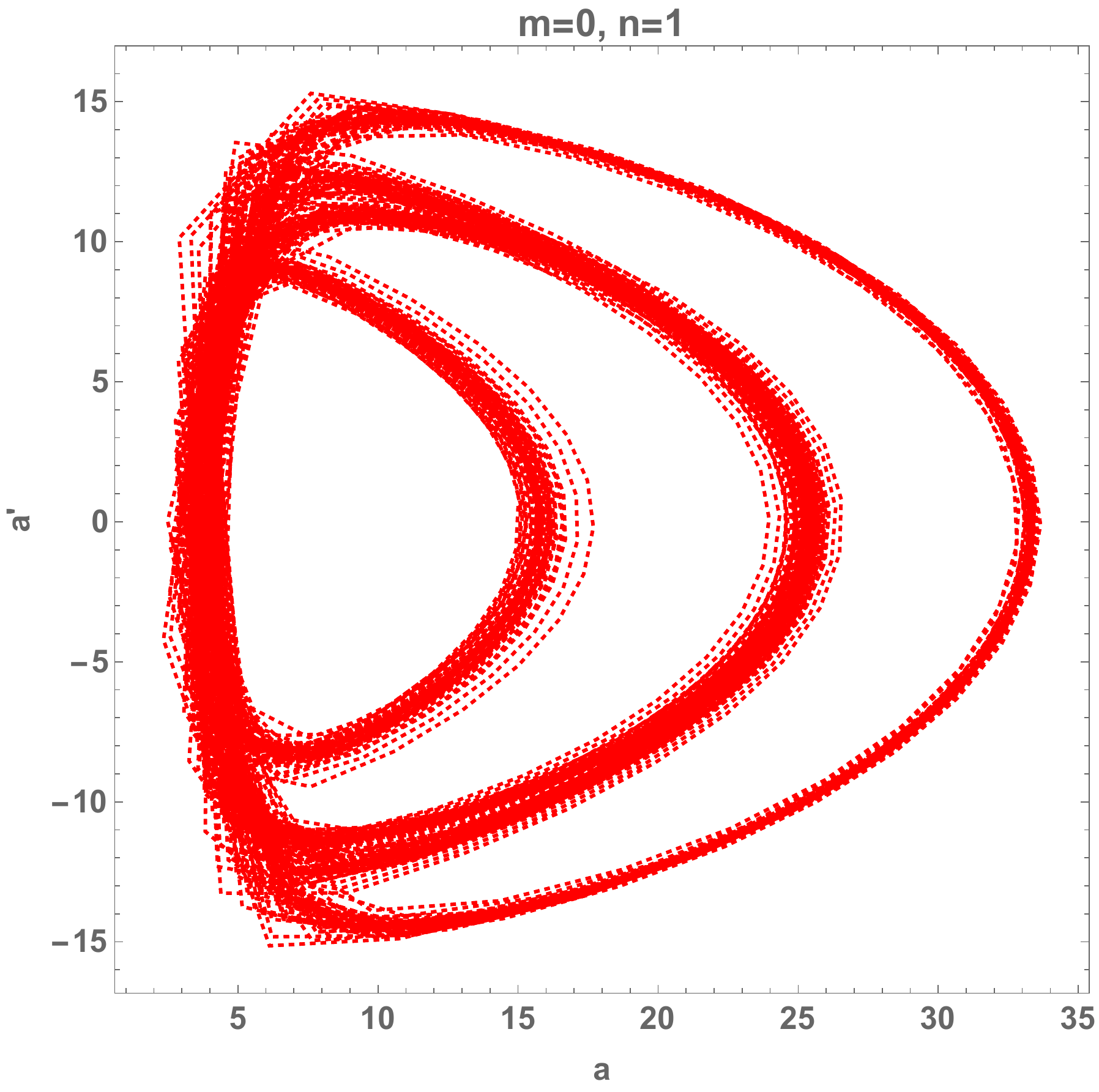}
\includegraphics[width= 0.45\columnwidth, height= 1.8 in]{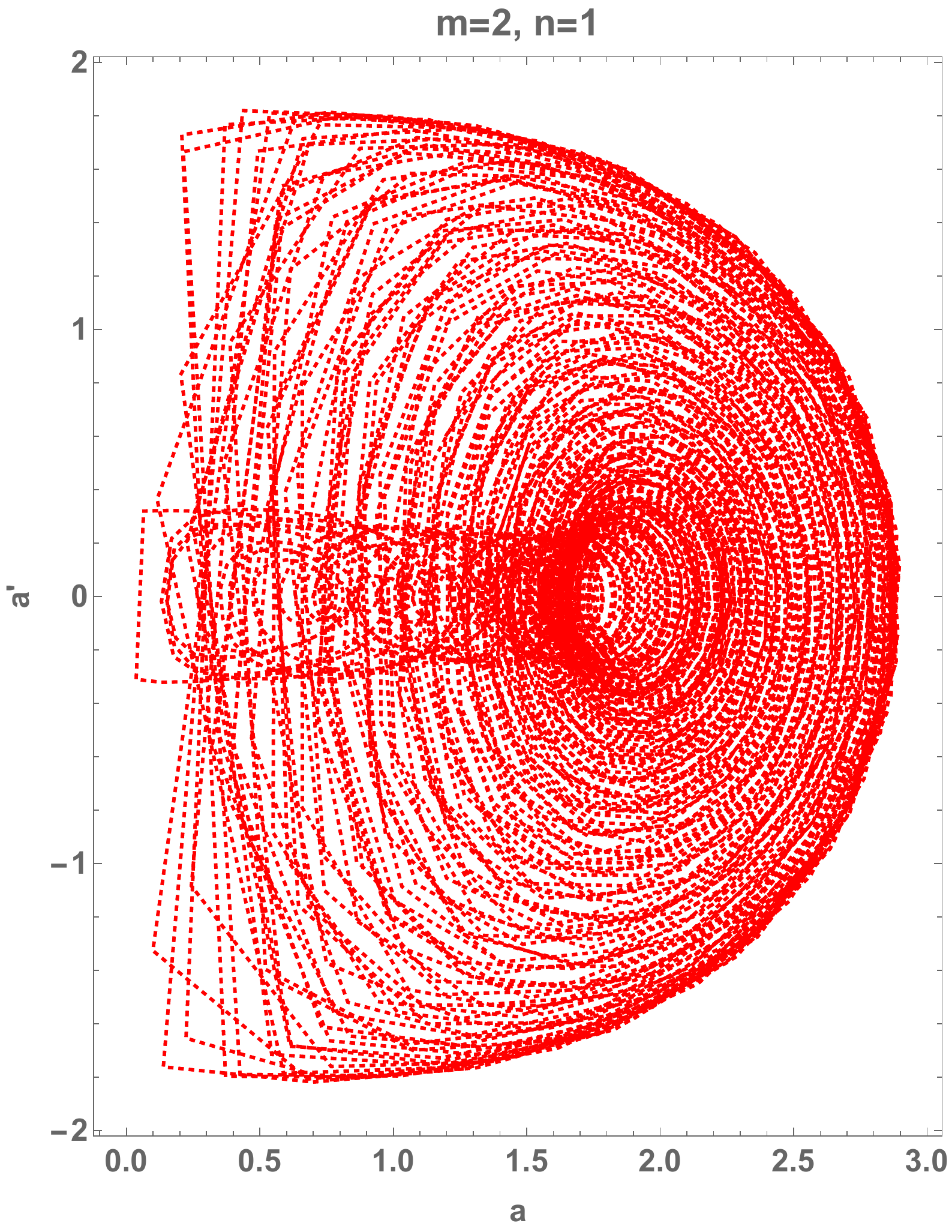} 
\includegraphics[width= 0.45\columnwidth, height= 1.8 in]{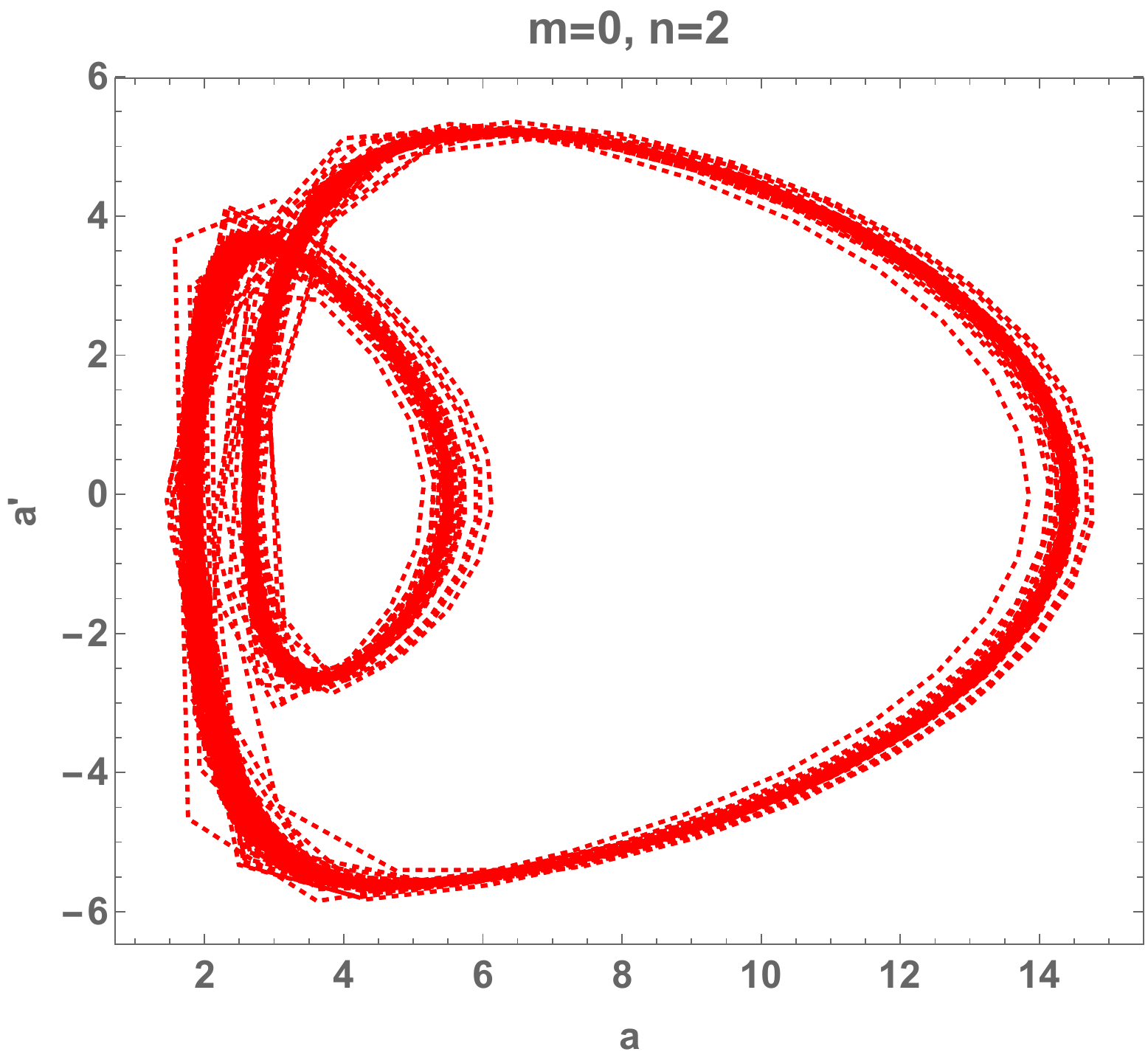} 
\includegraphics[width= 0.45\columnwidth, height= 1.8 in]{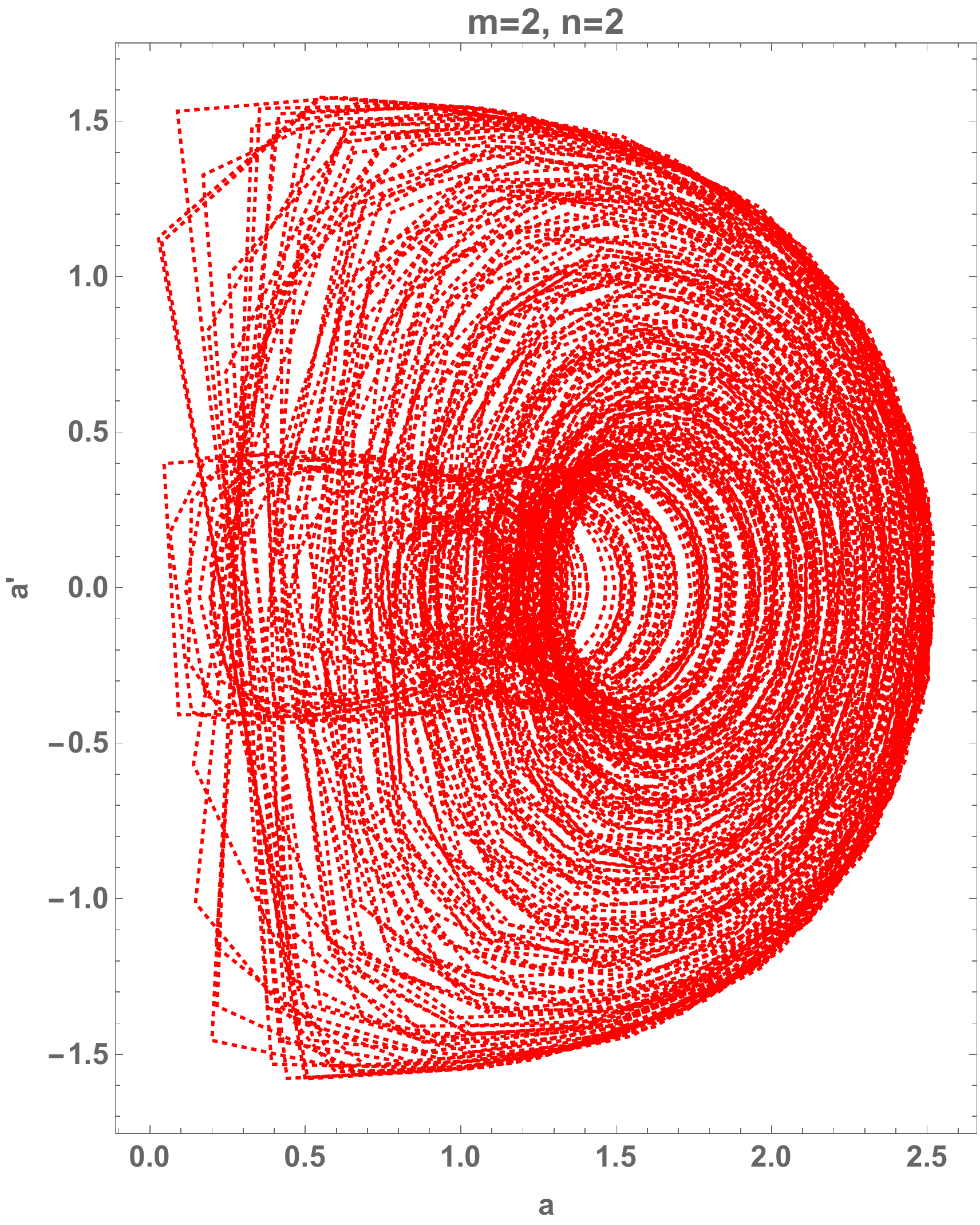} 
\caption{(Color online) Phase portrait of the laser amplitude $a(t)$, obtained from numerical simulations of eqs. (\ref{eq:foura})-(\ref{eq:fourd}). Parameter values are: $\omega=0.9$, $\theta=-0.7$ , $C=5$, $D=1$, $\gamma_r=0.7$, $\gamma_{im}=-0.14$, and $\Gamma=0.1$ for ($2,0$) and ($2,1$) media, and  $\gamma_{r}$=0.7, $\Gamma=0.02$ for ($2,2$). These parameters were varied near the values  $\omega=0.14$, $\theta=0.1$ , $C=1$, $D=1$, $\gamma_r=5$, $\gamma_{im}=-0.14$, and $\Gamma=0.04$ for ($2,0$) and ($2,1$) for ($0,0.5$), ($0,1$) and ($0,2$) pairs.}
\label{fig:foura}
\end{center}
\end{figure}

\begin{figure}[bt]
\begin{center}
\includegraphics[width= 0.45\columnwidth, height= 1.8 in]{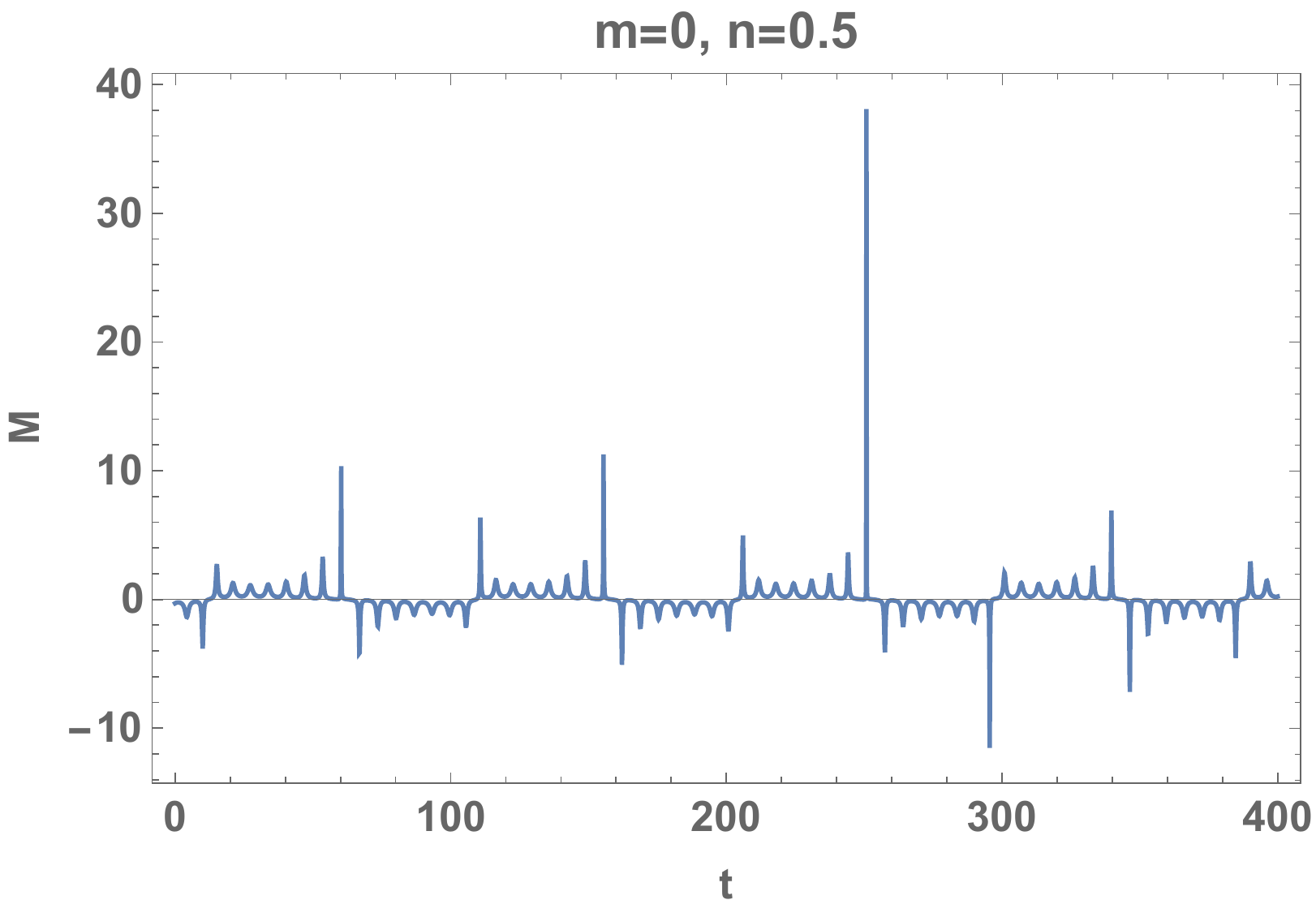}
\includegraphics[width= 0.45\columnwidth, height= 1.8 in]{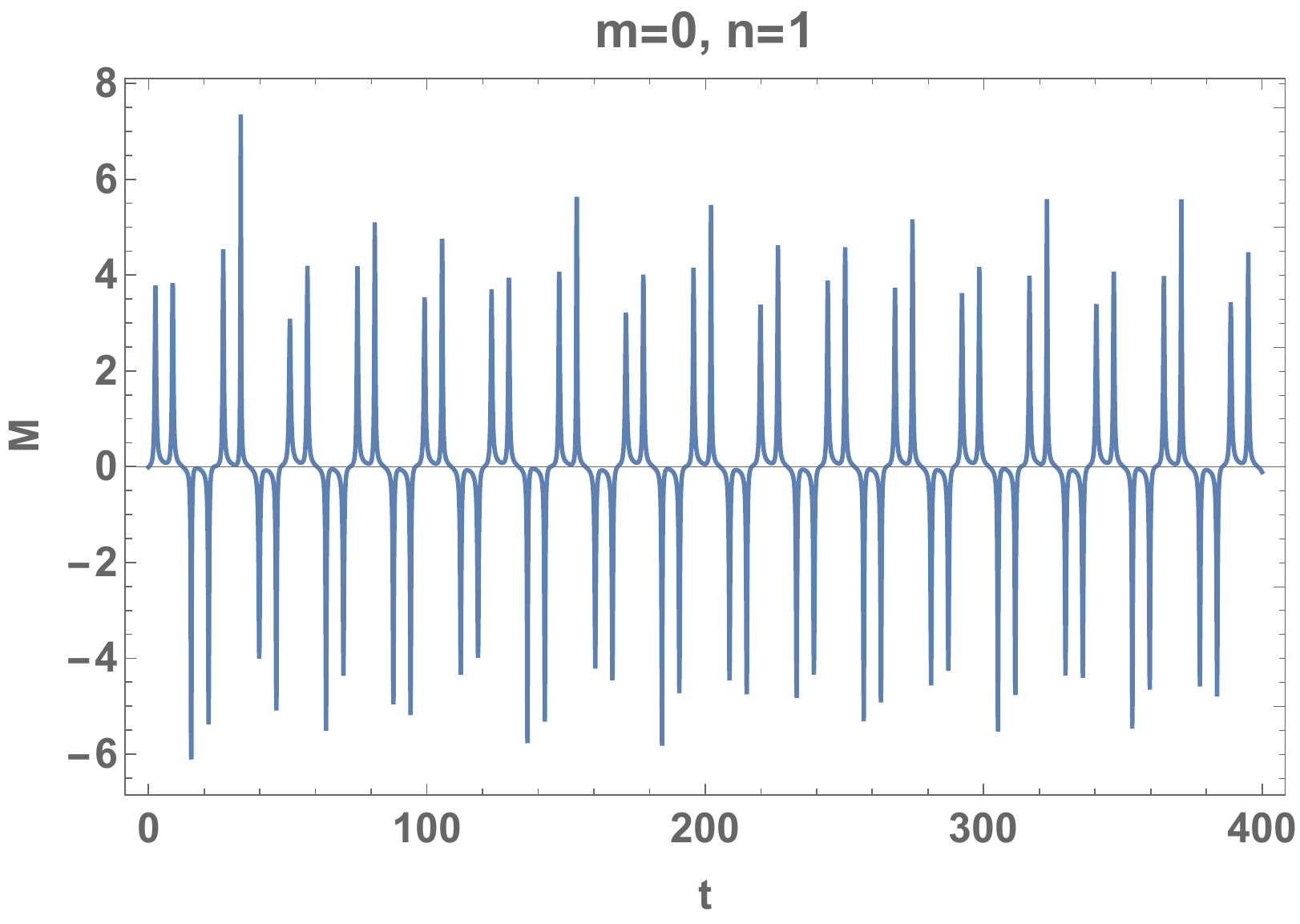}
\includegraphics[width= 0.45\columnwidth, height= 1.8 in]{multipulse01-Freq.pdf}
\includegraphics[width= 0.45\columnwidth, height= 1.8 in]{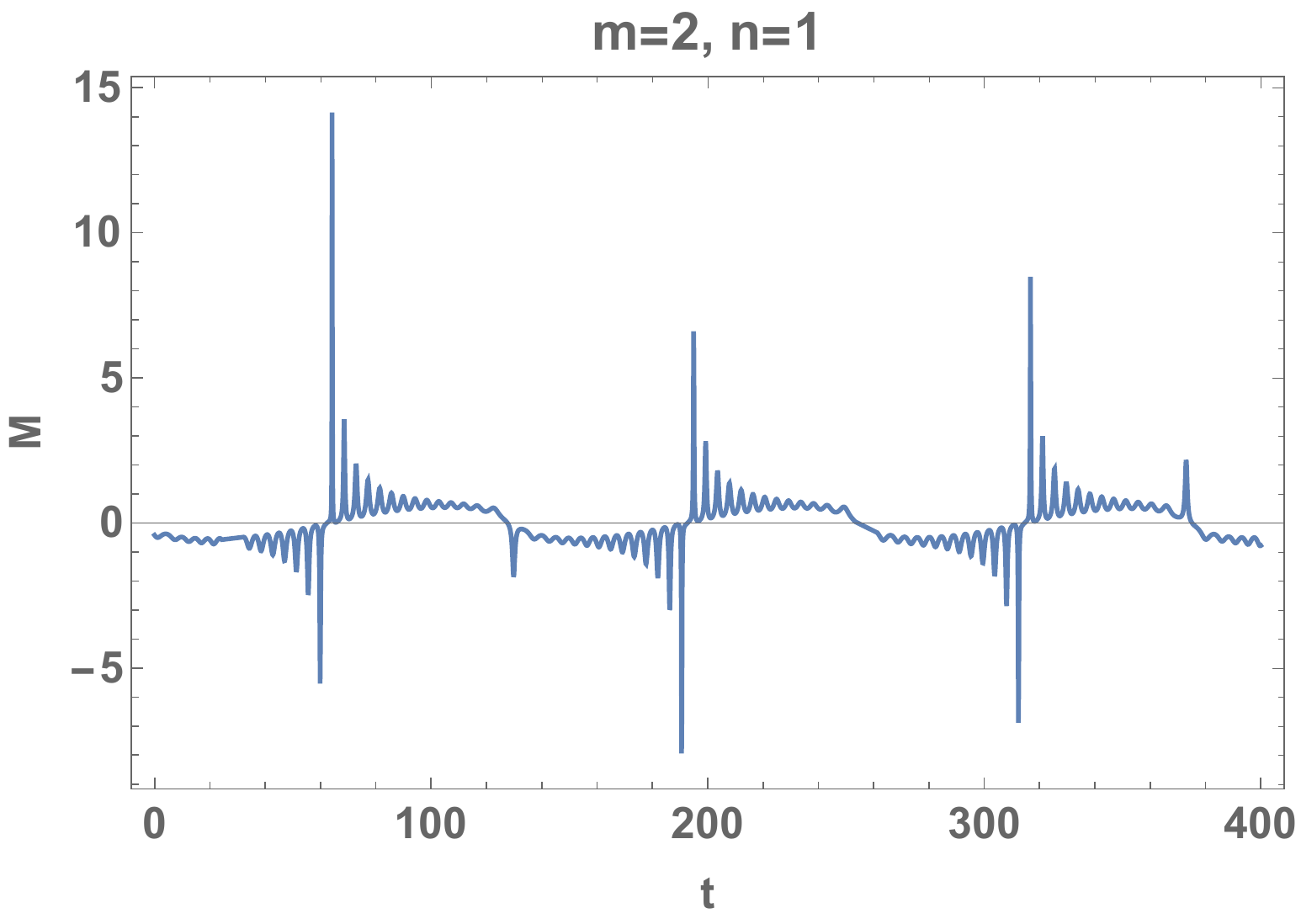} 
\includegraphics[width= 0.45\columnwidth, height= 1.8 in]{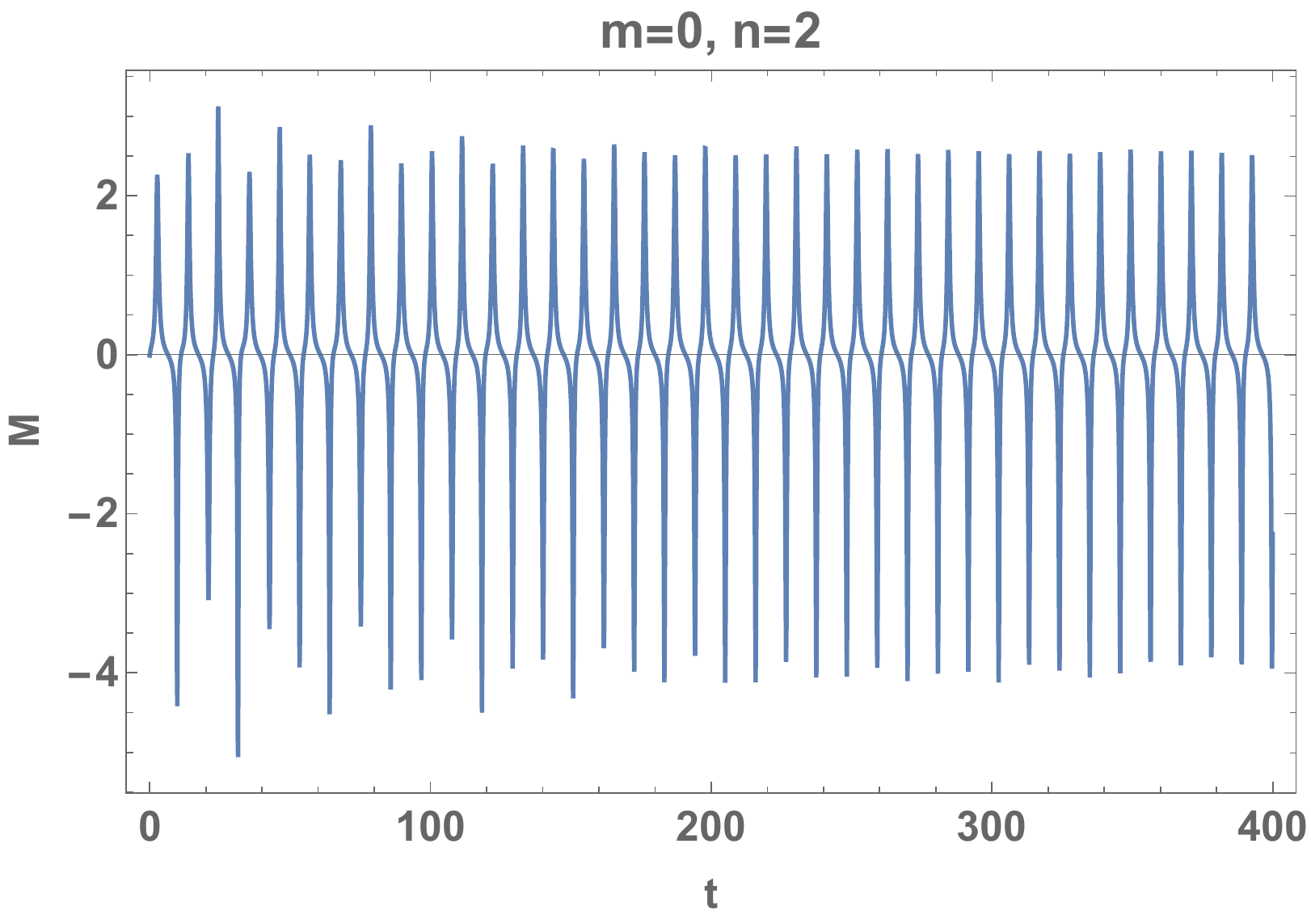} 
\includegraphics[width= 0.45\columnwidth, height= 1.8 in]{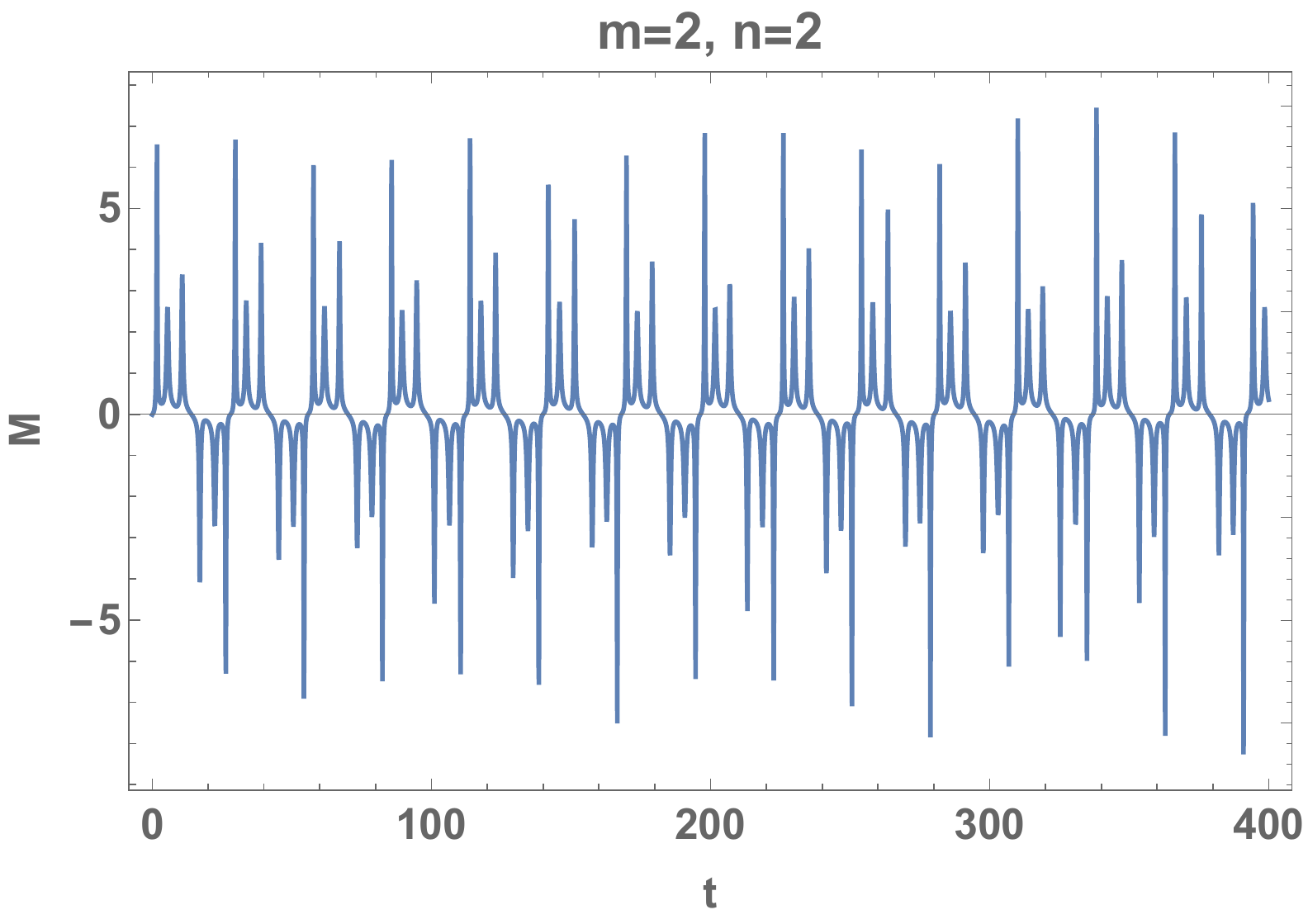} 
\caption{(color online) Time series of the instantaneous frequency $M(t)$, obtained from numerical simulations of eqs. \ref{eq:foura}-\ref{eq:fourd}. Numerical value of parameters $\omega$, $\theta$ , $C$, $D$, $\gamma_r$, $\gamma_{im}$, and $\Gamma=$ are the same as in Figs. \ref{fig:four} and \ref{fig:foura}.}
\label{fig:five}
\end{center}
\end{figure}

In the previous section we examined the stability of CWs, when their amplitude are coupled to a noise component and the packet grows in power to infinity as the real part of the noise amplification coefficient is positive. This growth, seen to be dependent on the noise modulation frequency $\Omega$, is reminiscent of CW instability therefore promoting laser self-starting. In the anomalous dispersion regime, the unstable CW field is expected to evolve into pulse trains or multi-pulse structures through the process of modulation instability \cite{Dikande2017,Mesumbe2019,a11,pd1}. \\
Profiles of the amplitude $a$ of pulse and multi-pulse structures for the present model, as well as the corresponding instantaneous frequency $M$, can be obtained by numerical simulations of eqs. (\ref{eq:foura})-(\ref{eq:fourd}). In this goal we employed a sixth-order Runge-Kutta algorithm with an adaptive step parameter \cite{luth}, to generate time series for the relevant dynamical parameters. Fig. \ref{fig:four} shows temporal profiles of the laser amplitude $a$, for some distinct combinations of the pair ($m,n$) and numerical values of parameters indicated in the captions.\\
For the pair ($0,0.5$), the laser amplitude is seen to deploy into a train of temporal pulses for which a large nonlinear gain ($\gamma_r$ = 4) is required to operate in multi-pulse mode, with the corresponding pulse peaks being reasonably high. The cases ($m,n$)=($2,0$) and ($2,1$) need not a high $\gamma_r$ value to operate into multi-pulse regime. However, even at high $\gamma_r$ (here $\gamma_r$=4), their peak amplitudes $a$ remain relatively small. For gain media with fast saturable absorber namely ($m,n$)=($0,1$) and ($0,2$), the multi-pulse regime settles only for large values of ($\gamma_r$), otherwise the laser will operate in CW mode. However this is not the case for the fiber laser with ($m,n$)=($2,2$), where a small value of $\gamma_r=0.7$ is enough to generate relatively high pulse trains. A global picture of time evolution of the laser amplitude $a$, for the various combinations ($m,n$) considered in Fig. \ref{fig:four}, is summarized in the phase-portrait representations shown in Fig. \ref{fig:foura}.\\
Fig. \ref{fig:five} shows time series of the instantaneous frequency $M(t)$, for different sets of values of the couple ($m,n$). 
Several windows of different average instantaneous frequencies are observed, each of these windows corresponds to a single pulse contained in the associated multi-pulse structures shown in Fig. \ref{fig:four}. Time series of $M(t)$ also reveals that widths of the bands in the periodic frequency-comb patterns are not equal, indeed pulses of wider bands are repeated at a small rate compared with narrower band pulses. Thus the multi-periodicity of the pulses could be useful for large band transmissions where for instance, the large bandwidth and low bandwidth pulses could be encoded with different carrier signals for long-distance communication purposes. Moreover, because the periods of the small bandwidth and large bandwidth pulses are quite different, the spectral filtering parameter $C$ can be used to tune multi-pulse structure either into a train of pulses (low-pass filtering) or into coupled ultrafast pulses (high-pass filtering). 

\section{Conclusion}\label{section7}
We have studied the dynamics and stability of CW and multi-pulse structures in a generalized model of complex Ginzburg-Landau equation, aimed to describe the propagation of optical fields in a mode-locked fiber laser with an arbitrary nonlinearity. The model encompasses a broad range of physical contexts characterized by distinct strengths of nonlinearity of the optical medium, ranging from cubic nonlinearity to saturable-nonlinearity fiber lasers via systems with cubic-quintic nonlinearity. Results of the stationary CW solutions for the model have shown that there exists zero, one or many fixed points in the a-M plane for nonlinearity saturation with $n>1$, therefore confirming the role of saturation as a mean to sustain the CW solutions. The stability analysis of CWs solutions, via the modulational-instability approach, indicated a bifurcation of the noise growth rate and propagation constant within a specific range of the noise modulation frequency. This effect was associated with the propagation of two perturbation fields that can evolve into multi-pulses structures. Numerical simulations of the optical field amplitude and its instantaneous modulation frequency, unveiled the structures of multi-pulse patterns for few illustrative values of the couple ($m,n$) characterizing the nonlinearity-saturation law. Our results are consistent with previous works achieved on known fiber lasing systems. \\
Although the present study was intended for passively mode-locked fiber lasers with saturable absorber, the generalized model can very well be suitably applied to advanced optical fiber materials that exhibit high-order (i.e. non-Kerr) optical nonlinearities \cite{pd1}.

\section*{Conflict of interest}
The authors declare no conflict of interest.

\section*{Acknowledgments}

A. M. Dikand\'e thanks Holger Kantz, Head of research unit "Nonlinear Dynamics and Time-Series analysis" at the Max Planck Institute for the Physics of Complex Systems (MPIPKS), Dresden, Germany, for permitting a visit during which part of this work was done. 


\begin{thebibliography}{99}

\bibitem{1} J. Herrmann, V. P. Kalosha, and M. M\"uller, {\it Higher-order phase dispersion in femtosecond Kerr-lens mode-locked solid-state lasers: sideband generation and pulse splitting}, Opt. Lett. \textbf{22}, 236 (1997).
\bibitem{2} A. I. Chernykh and S. K. Turitsyn, {\it Soliton and collapse regimes of pulse generation in passively mode-locking laser systems}, Opt. Lett. \textbf{20}, 398 (1995).
\bibitem{3} K. M. Spaulding, D. H. Yong, A. D. Kim, and J. N. Kutz, {\it Nonlinear dynamics of mode-locking optical fiber ring lasers}, J.
Opt. Soc. Am. B\textbf{19}, 1045 (2002).
\bibitem{4} M. E. Fermann, {\it Ultrashort-pulse sources based on single-mode rare-earth-doped fibers}, J. Appl. Phys. B \textbf{58}, 197 {1994}.
\bibitem{33} H. A. Haus, {\it Mode-locking of lasers}, IEEE J. Select. Topics Quantum Elect. \textbf{6}, 1173 (2000).
\bibitem{44} U. Keller, D. Miller, G. Boyd, T. Chiu, J. Ferguson, and M. Asorn,
{\it Solid-state low-loss intracavity saturable absorber for Nd:YLF lasers:
An antiresonant semiconductor Fabry-Perot saturable absorber}, Opt.
Lett. \textbf{17}, 505 (1992).
\bibitem{66} V. J. Matsas, T. P. Newson, D. J. Richardson, and D. J. Payne, {\it Self-starting passively mode-locked fiber ring soliton laser exploiting non
linear polarization rotation}, Electron. Lett. \textbf{28}, 1391 (1992).
\bibitem{77} N. J. Doran and D. Wood, {\it Nonlinear optical loop mirror}, Opt. Lett. \textbf{14}, 56 (1988).
\bibitem{99} W. S. Wong, S. Namiki, M. Margalit, H. A. Haus, and E. P. Ippen, {\it Self-switching of optical pulses in dispersion-imbalanced nonlinear loop mirrors}, Opt. Lett. \textbf{22}, 1150 (1997).  
\bibitem{a1} H. A. Haus, {\it Theory of mode locking with a fast saturable absorber }, J. Appl. Phys. 46, 3049 (1975).
\bibitem{a2} H. A. Haus and Y. Silberberg, {\it Laser mode locking with addition of nonlinear index}, IEEE J. Quantum Elec. \textbf{22}, 325 (1986).
\bibitem{a3} E. P. Ippen, H. A. Haus and L. Y. Liu, {\it Additive pulse mode locking }, J. Opt. Soc. Am. B\textbf{6}, 1736 (1989).
\bibitem{a4} O. E. Martinez, R. L. Fork and J. P. Gordon, {\it Theory of passively mode-locked lasers including self-phase modulation and group-velocity dispersion}, Opt. Lett. \textbf{9}, 156 (1984).
\bibitem{a6} C. J. Chen, P. K. A. Wai and C. R. Menyuk, {\it Self-starting of passively mode-locked lasers with fast saturable absorbers}, Opt. Lett.
\textbf{20}, 350 (1994).
\bibitem{a8} V. L. Kalashnikov, E. Sorokin and I. T. Sorokina, {\it Multipulse operation and limits of the Kerr-lens mode-locking stability}, IEEE
J. Quant. Elec. \textbf{39}, 323 (2003).
\bibitem{a9} H. A. Haus, E. P. Ippen and K. Tamura, {\it Additive-pulse modelocking in fiber lasers}, IEEE J. Quan-
tum Electron. \textbf{30}, 200 (1994).
\bibitem{a10} J. M. Soto-Crespo and N. N. Akhmediev, {\it Multisoliton regime of pulse generation by lasers passively mode locked with a slow saturable absorber}, J. Opt. Soc.
Am. B\textbf{16}, 674 (1999).
\bibitem{p1}V. V. Afanasjev, B. A. Malomed and P. L. Chu, {\it Stability of bound states of pulses in the Ginzburg-Landau equations}, Phys. Rev. E\textbf{56}, 6020 (1997).
\bibitem{p2}M. Olivier and M. Pich\'e, {\it Origin of the bound states of pulses in the stretched-pulse fiber laser}, Opt. Expr. \textbf{17}, 405 (2009).
\bibitem{p3} L. Yun and X. Liu, {\it Generation and Propagation of Bound-State Pulses in a Passively Mode-Locked Figure-Eight Laser}, IEEE Photon. J. \textbf{4}, 512 (2012).
\bibitem{a11} J. M. Soto-Crespo, N. N. Akhmediev and G. Town, {\it Continuous-wave versus pulse regime in a passively mode-locked laser with a fast saturable
  absorber}, J. Opt. Soc. Am. B\textbf{19}, 234 (2002).11
\bibitem{a12} J. M. Soto-Crespo, M. Grapinet, P. Grelu and N. N. Akhmediev, {\it Bifurcations and multiple-period soliton pulsations in a passively mode-locked fiber laser}, Phys. Rev. E\textbf{70}, 066612 (2004).
\bibitem{a13} J. W. Haus, M. Hayduk, W. Kaechele, G. Shaulov, J.
Theimer, K. Teegarden and G. Wicks, {\it A mode-locked fiber laser with a chirped grating mirror},  Opt. Commun. \textbf{174}, 205 (2000). 
\bibitem{pd1} J. M. Hickmann, S. B. Cavalcanti, N. M. Borges, E. A.
Gouveia and A. S. Gouveia-Neto, {\it Modulational instability in semiconductor-doped glass fibers with saturable nonlinearity}, Opt. Lett. \textbf{18}, 182 (1993).
\bibitem{pd2} M. L. Lyra and A. S. Gouveia-Neto, {\it Saturation effects on modulational instability in non-Kerr-like monomode optical fibers}, Opt. Commun. \textbf{108}, 117 (1994).
\bibitem{Dikande2017}
A.~M. Dikand\'e, J. Voma Titafan and B.~Z. Essimbi, {\it Continuous-wave
  to pulse regimes for a family of passively mode-locked lasers with saturable
  nonlinearity,} J. Opt. \textbf{19}, 105504 (2017).
\bibitem{Mikhail2014} M.~Lapine, I.~V. Shadrivov, and Y.~S. Kivshar, {\it Colloquium: Nonlinear
  metamaterials,} Rev. Mod. Phys. \textbf{86}, 1093--1123 (2014).
\bibitem{Petmegni2017}D.~Petmegni Mbieda, A. M. Dikandé and B.~Essimbi, {\it Raman
  self-induced-transparency soliton trains in hollow-core photonic crystals},
  Applied Physics B \textbf{123}, 171 (2017).  
\bibitem{Huang2016}
Y.-Q. Huang, Z.-A. Hu, H.~Cui, Z.-C. Luo, A.-P. Luo, and W.-C. Xu,
  {\it Coexistence of harmonic soliton molecules and rectangular noise-like
  pulses in a figure-eight fiber laser}, Opt. Lett. \textbf{41}, 4056 (2016).
\bibitem{Mesumbe2019}
E.~Mesumbe and A. M.~Dikand\'e, {\it Modulational instability and soliton
  trains in a model for two-mode fiber ring lasers}, Opt. Quant.
  Elect. \textbf{51}, 361 (2019).
\bibitem{Dikande2010}
A.~M. Dikand\'e, {\it Fundamental modes of a trapped probe photon in
  optical fibers conveying periodic pulse trains}, Phys. Rev. A \textbf{81},
  013821 (2010).  
\bibitem{Dikande2011}
A.~M. Dikand\'e, {\it Induced soliton ejection from a continuous-wave
  source waveguided by an optical pulse soliton train}, J. Opt.
  \textbf{13}, 035203 (2011).
\bibitem{MbiedaPetmegni2017}
D.~Petmegni Mbieda and A. M.~Dikand\'e, {\it Periodic soliton trains in
  nonlinear magneto-optical media,} J. Mod. Opt. \textbf{64},
  1192 (2017). 
  \bibitem{brice}A. B. Moubissi, Th. B. Ekogo, S. D. Bidouba Sanvany, Z. H. Moussambi Membetsi and A. M. Dikand\'e, {\it Averaged-dispersion management for ultrashort soliton molecule propagation in lossy fibre systems}, Opt. Commun. \textbf{431}, 187 (2019).
\bibitem{Amrani2011}
F.~Amrani, A.~Niang, M.~Salhi, A.~Komarov, H.~Leblond, and F.~Sanchez,
  {\it Passive harmonic-mode locking of soliton crystals}, Opt. Lett.
  \textbf{36}, 4239 (2011).
\bibitem{Jubgang2015}
D.~J. Fandio~Jubgang, A.~M. Dikand\'e, and A.~Sunda-Meya, {\it Elliptic
  solitons in optical fiber media}, Phys. Rev. A \textbf{92}, 053850 (2015).
\bibitem{welak}D. D. M. Welakuh and A. M. Dikand\'e, {\it Storage and retrieval of time-entangled soliton trains in a three-level atom system coupled to an optical cavity}, Opt. Commun. \textbf{403}, 27 (2017).
\bibitem{fandioa}D.~J. Fandio~Jubgang and A.~M. Dikand\'e, {\it Pulse train uniformity and nonlinear dynamics of
soliton crystals in mode-locked fiber ring lasers}, J. Opt. Soc. Am. B\textbf{37}, 2721 (2017).
\bibitem{DikandeBitha2019}
R.~Dikand\'e Bitha and A. M. Dikand\'e, {\it Soliton-comb structures in ring-shaped optical microresonators: generation, reconstruction and stability}, Eur. Phys. J. D \textbf{73}, 152 (2019).
\bibitem{DikandeBitha2019a} R.~Dikand\'e~Bitha and A. M.
~Dikand\'e, {\it Elliptic-type soliton combs in optical ring microresonators}, Phys. Rev. A \textbf{97}, 033813 (2018).
\bibitem{luth}H. Luther, {\it An Explicit Sixth-Order Runge-Kutta Formula}, Math. Comp. {\bf 22} 434 (1968).
\bibitem{FMbieda2020}
F.~G.~Ngomegni Mbieda, A.~M. Dikand\'e, and B.~Z. Essimbi, {\it Dynamics
  and stability of cw and pulse lasers in kerr optical media with K-photon
  absorption}, Phys. Scrip. \textbf{95}, 025502 (2020).
\bibitem{kam}P. Kameni Nteutse, A. M. Dikand\'e and S. Zekeng, {\it Competing effects of Kerr nonlinearity and K-photon absorptions on continuous-wave laser inscriptions}, Opt. Quantum Elec. \textbf{52}, 313 (2019).
\end{thebibliography}

\end{document}